%% file: eseSpectra.tex
\newif\ifdraft
\newif\iffull
\newif\ifcomment
\newif\ifpreprint
\preprinttrue
\fulltrue
\def\dvers{v3.0}
\ifpreprint
\documentclass[ALICE,manyauthors]{cernphprep}
\else
\documentclass[10pt,aps,prl,twocolumn,superscriptaddress,altaffilletter,nobibnotes,nofootinbib]{revtex4-1}
\fi

\usepackage{svn-multi}
\svnid{$Id: eseSpectra.tex 1065 2015-06-23 15:29:14Z lmilano $}
\usepackage{graphicx}  
\usepackage{bm}        
\usepackage{amssymb}   
\usepackage{amsfonts}
\usepackage{graphics}
\usepackage{epsfig}
\usepackage[normalem]{ulem}
\usepackage[usenames]{color}
\usepackage{multirow}
\usepackage{units}
\usepackage{booktabs}
\usepackage{tabularx}
\usepackage{hyperref}
\RequirePackage{color}\definecolor{RED}{rgb}{1,0,0}\definecolor{BLUE}{rgb}{0,0,1}

\input{commands.tex}
\graphicspath{{./img/}}
\ifdraft
\usepackage{lineno}
\linenumbers
\modulolinenumbers[5]
\usepackage{fancyhdr}
\else
\renewcommand{\warn}[1]{}

\fi
\renewcommand{\warn}[1]{} 
\begin{document}
\ifpreprint
\begin{titlepage}
\PHyear{2015}
\PHnumber{171}                 
\PHdate{01 July}              
%
%
\title{Event shape engineering for inclusive spectra and elliptic flow\\in Pb--Pb collisions at \snnbf~=~2.76 TeV}
\ShortTitle{Event shape engineering in Pb--Pb collisions at \snn~=~2.76 TeV} 

%
\Collaboration{ALICE Collaboration%
         \thanks{See Appendix~\ref{app:collab} for the list of collaboration
                      members}}
\ShortAuthor{ALICE Collaboration}      
\begin{abstract}
\input{abstract.tex}
\end{abstract}
\end{titlepage}
\else
\title{Multiplicity Dependence of Pion, Kaon, Proton and Lambda Production in p--Pb Collisions at \snn~=~5.02~TeV}

\ifdraft
\date{\today, \color{red}DRAFT \dvers\ \$Revision: 989 $\color{white}:$\$\color{black}}
\else
\date{\today}
\fi
\begin{abstract}
\input{abstract.tex}
\end{abstract}
\maketitle
\fi
\ifdraft
\thispagestyle{fancyplain}
\fi
\setcounter{page}{2}
\input{eseSpectraMain.tex}

\iffull
\section*{Acknowledgments}
\input{acknowledgements.tex}
\fi

\bibliographystyle{utphys}
\bibliography{eseSpectra}

\iffull
\newpage
\appendix
\section{The ALICE Collaboration}
\label{app:collab}
\input{Alice_Authorlist_2015-Jun-23.tex}

\fi

\end{document}

%% file: commands.tex
\newcommand{\TPC}          {\rm{TPC}}
\newcommand{\VZERO}        {\rm{V0}}
\newcommand{\VZEROA}       {\rm{V0A}}
\newcommand{\VZEROC}       {\rm{V0C}}

\newcommand{\pbar}         {$\rm\overline{p}$}

\newcommand{\dedx}         {\ensuremath{\mathrm{d}E/\mathrm{d}x}}

\newcommand{\pp}           {pp}

\newcommand{\PbPb}         {\mbox{Pb--Pb}}
\newcommand{\pPb}          {\mbox{p--Pb}}

\newcommand{\nsigma}       {\ensuremath{n\sigma}}
\newcommand{\dcaxy}        {\ensuremath{{\rm DCA}_{xy}}}
\newcommand{\dcaz}         {\ensuremath{{\rm DCA}_{z}}}

\newcommand{\pt}           {\ensuremath{p_{\rm T}}}

\newcommand{\snn}          {\ensuremath{\sqrt{s_{\rm NN}}}}
\newcommand{\snnbf}        {\ensuremath{\mathbf{{\sqrt{\textit{s}_{\mathbf NN}}}}}}

\newcommand{\avbT}         {\ensuremath{\left< \beta_{\rm T}\right>}}
\newcommand{\avpT}         {\ensuremath{\left< \pt \right>}}

\newcommand{\Fig}[1]       {Fig.~\ref{#1}}
\newcommand{\Figure}[1]    {Figure~\ref{#1}}
\newcommand{\Sect}[1]      {Sect.~\ref{#1}}
\newcommand{\Section}[1]   {Section~\ref{#1}}

\newcommand{\gevc}         {\ensuremath{{\rm GeV}/c}}

\newcommand{\vn}           {\ensuremath{v_{n}}}
\newcommand{\vtwo}         {\ensuremath{v_{2}}}
\newcommand{\vtsp}         {\ensuremath{v_{2}\{\mathrm{SP}\}}}
\newcommand{\qtwo}         {\ensuremath{q_{2}}}

\newcommand{\qtvc}         {\ensuremath{q_{2}^{\VZEROC}}}
\newcommand{\qttpc}        {\ensuremath{q_{2}^{\rm TPC}}}
\newcommand{\qtrec}        {\ensuremath{q_{2}^{\rm rec}}}
\newcommand{\qtgen}        {\ensuremath{q_{2}^{\rm gen}}}
\newcommand{\hq}           {\ensuremath{{\rm large-}q_{2}}}
\newcommand{\sq}           {\ensuremath{{\rm small-}q_{2}}}
\newcommand{\hqnoq}           {\ensuremath{{\rm large-}}}

\newcommand{\vthree}       {\ensuremath{v_{3}}}
\newcommand{\um}           {\ensuremath{\mu{\rm m}}}

\newcommand{\Deta}         {\ensuremath{\Delta\eta}}
\newcommand{\zvtx}         {\ensuremath{z_{\mathrm{vtx}}}}

\newcommand{\warn}[1]      {{\small\textbf{\textcolor{red}{(!\footnote{\textbf{(!)}~#1})}}}}



%% file: abstract.tex
We report on results obtained with the Event Shape Engineering
technique applied to Pb--Pb collisions at \snn~=~2.76~TeV.  By
selecting events in the same centrality interval, but with very
different average flow, different initial state conditions can be
studied.  We find the effect of the event-shape selection on the
elliptic flow coefficient $v_2$ to be almost independent of transverse
momentum \pt, as expected if this effect is due to fluctuations in the
initial geometry of the system. Charged hadron, pion, kaon, and proton
transverse momentum distributions are found to be harder in events
with higher-than-average elliptic flow, indicating an interplay
between radial and elliptic flow.

%% file: eseSpectraMain.tex
\section{Introduction}
\label{sec:introduction}

Results from Lattice Quantum
Chromo-Dynamics~\cite{Borsanyi:2010bp,Bazavov:2011nk} predict the
existence of a plasma of deconfined quarks and gluons, known as the
``Quark Gluon Plasma'' (QGP).  This state of matter can be produced in
the laboratory by colliding heavy nuclei at relativistic
energies~\cite{Abreu:2007kv,Schukraft:2013wba,Akiba:2015jwa}.  The QGP
was found to behave as a nearly perfect liquid and its properties can
be described using relativistic hydrodynamics (for a recent review,
see~\cite{Gale:2013da}). The current experimental heavy-ion programs
at Brookhaven's Relativistic Heavy Ion Collider (RHIC) and at CERN's
Large Hadron Collider (LHC) are aimed at a precise characterization of
the QGP, in particular of its transport properties.

The system created in a heavy-ion collision expands and hence cools
down, ultimately undergoing a phase transition to a hadron gas, which
then decouples to the free-streaming particles detected in the
experiments~\cite{Gale:2013da}. A precision study of the QGP
properties requires a detailed understanding of this expansion
process.  If the initial geometry of the interaction region is not
azimuthally symmetric, a hydrodynamic evolution of a nearly-ideal
liquid (i.e. with a small value of the shear viscosity over entropy
ratio $\eta/s$) gives rise to an azimuthally anisotropic distribution
in momentum space for the produced particles. This anisotropy can be
characterized in terms of the Fourier coefficients \vn\ of the
particle azimuthal distribution~\cite{Heinz:2013th}. The shape of the
azimuthal distribution, and hence the values of these Fourier
coefficients, depend on the initial conditions and on the expansion
dynamics. 
The geometry of the initial state fluctuates event-by-event and
measurements of the resulting $v_n$ fluctuations pose stringent
constraints on initial state models.  A quantitative understanding of
the initial geometry of the produced system is therefore of primary
importance~\cite{Gale:2013da}.  A number of different experimental
measurements and techniques have been proposed to disentangle the
effects of the initial conditions from QGP transport, including
measurements of correlations of different
harmonics~\cite{Aad:2014fla}, event-by-event flow
fluctuations~\cite{Alver:2006wh,Alver:2007qw,Aad:2013xma,Abelev:2012di}
and studies in ultra-central
collisions~\cite{ALICE:2011ab,CMS:2013bza}.  Recent results from
\pp\ and \pPb\ collisions at the LHC, moreover, suggest that
hydrodynamic models may be also applicable to small
systems~\cite{ABELEV:2013wsa,Abelev:2013haa,Werner:2013ipa,Shuryak:2013ke,Bozek:2013yfa}.
This further highlights the importance of studying \PbPb\ collisions
with more differential probes, to investigate the interplay between
the initial conditions and the evolution, in the system where the
hydrodynamic models are expected to be most applicable.
 
One of the new tools for the study of the dynamics of heavy-ion
collisions is the ``Event Shape Engineering''
(ESE)~\cite{Schukraft:2012ah}. This technique is based on the
observation that the event-by-event variation of the anisotropic flow
coefficient (\vn) at fixed centrality is very
large~\cite{Abelev:2012di}.  Hydrodynamic calculations show that the
response of the system to the initial spatial anisotropy is
essentially linear for the second and third harmonic, meaning that the
final state \vtwo\ (and \vthree) are very well correlated with the
second (and third) order eccentricities in the initial state for small
values of $\eta/s$~\cite{Heinz:2013th,Gardim:2012dc,Voloshin:2008dg}.
These observations suggest a possibility to select events in heavy-ion
collisions based on the initial (geometrical) shape, providing new
opportunities to study the dynamics of the system evolution and the
role of the initial conditions.

The ESE technique is proposed to study ensemble-averaged observables
(such as \vtwo\ and inclusive particle spectra) in a class of events
corresponding to the same collision centrality, but different
\vn\ values. In this paper events are selected based on the magnitude
of the second order reduced flow vector \qtwo~(see
\Sect{sec:event-selection}).  The technique was recently applied to
study correlations between different flow harmonics in the ATLAS
experiment~\cite{Aad:2015lwa}.  In this paper we present the results
on elliptic flow and charged particle specta in \PbPb\ collisions at
\snn~=~2.76~TeV obtained with ESE technique.  The events selected with
the ESE technique are characterized by the measurement of \vtwo, to
quantify the effect of the selection on the global properties of the
event.  In order to search for a connection between elliptic and
radial flow the effect of the ESE selection on the inclusive
transverse momentum distribution of charged hadrons, pions, kaons and
protons is then studied.  The results are presented for primary
charged particles, defined as all prompt particles produced in the
collision including all decay products, except those from weak decays
of light flavor hadrons and of muons. The differential measurement
described in this work could provide important constraints to identify
the correct model for initial conditions and for the determination of
transport properties. The development of flow in hydrodynamical
models is driven by the pressure gradients and anisotropy
in the initial state. A correlation between anisotropic and radial
flow may stem from the specific fluctuation pattern in the initial
state and/or can be produced in the final state depending on the bulk
and shear viscosity of the system~\cite{Heinz:2013th}.

A few important caveats, which can affect the selectivity of the ESE
technique, have to be kept in mind in this study.  First, the
discriminating power of the \qtwo\ selection depends on the
multiplicity and \vtwo\ value in the psudorapidity, $\eta$, region
where it is computed and on the intrinsic resolution of the detector
used for the measurement.  Second, non-flow effects (such as resonance
decays, jets, etc.~\cite{Voloshin:2008dg}) could bias the
\qtwo\ measurement.  In this work we discuss both aspects in detail,
making use of different detectors with different intrinsic resolution
and different $\eta$ coverage.

The paper is organized as follows. In \Sect{sec:alice-detector-data} a
brief review of the ALICE detector and of the data sample is
presented. In \Sect{sec:analysis-technique} the analysis technique,
with an emphasis on the event selection and on the particle
identification strategy, is discussed. The results are presented in
\Sect{sec:results}. Their implication for the hydrodynamic
interpretation is discussed in \Sect{sec:discussion}. Finally, we come
to our conclusions in \Sect{sec:conclusion}.

\section{ALICE detector and data sample}
\label{sec:alice-detector-data}                  

The ALICE detector at the CERN LHC was designed to study mainly
high-energy \PbPb\ collisions.  It is composed of a central barrel
($\left|\eta\right| \lesssim 0.8$ for full-length tracks), containing
the main tracking and particle identification detectors, complemented
by forward detectors for specific purposes (trigger, multiplicity
measurement, centrality determination, muon tracking). A detailed
description of the apparatus can be found in~\cite{Aamodt:2008zz}. The
main detectors used for the analysis presented in this paper are
discussed below.

The main tracking devices in the central barrel are the Inner Tracking
System (ITS) and the Time Projection Chamber (TPC). They are immersed
in a 0.5~T solenoidal field. The ITS is the detector closest to the
interaction point. It is a six-layer silicon tracker with a very low
material budget ($\sim 7\%$ of one radiation length $X_0$).  The ITS
provides information on the primary interaction vertex and is used to
track particles close to the interaction point, with the first layer
positioned at a radial distance of 3.9~cm from the interaction point
and the sixth one at 43~cm. It can measure the transverse impact
parameter (\dcaxy) of tracks with a resolution of about 300 (40)~\um,
for transverse momentum $\pt = 0.1\ (4)~\gevc$, allowing the
contamination from secondary particles to be significantly
reduced. The TPC~\cite{Alme:2010ke} is a large-volume gas detector
(external diameter 5~m) which measures up to 159 space points per
track, providing excellent tracking performance and momentum
resolution ($\sigma_{\pt}/\pt\sim 6\%$ at $\pt =
10~\gevc$)~\cite{Abelev:2014ffa}. It is also used in this work to
identify particles through the measurement of the specific energy
loss, \dedx. The \dedx, computed as a truncated mean utilizing only
60\% of the available samples, has a resolution
of $\sim$~5\% in peripheral and $\sim$~6.5\% in central collisions~\cite{Abelev:2014ffa}. At
a radius of 3.7~m from the beam axis, the Time of Flight (TOF)
detector measures the arrival time of particles with a total
resolution of about 85~ps in \PbPb\ collisions, allowing a $\pi$/K
(K/p) 2~$\sigma$ separation up to $\pt = 3~(5)~ \gevc$.  The ALICE
reconstruction software performs tracking based either on the
information from the TPC alone (TPC-only tracks) or on the combined
information from the ITS and TPC (global tracks). The former have the
advantage of an essentially flat azimuthal acceptance, and are used
for \vtwo\ and \qtwo\ measurements. The latter provide better quality
tracks ($\sigma_{\pt}/\pt \sim 1.5\%$ at $\pt =
10~\gevc$)~\cite{Abelev:2014ffa}, rejecting most of the secondary
tracks.  However, the acceptance and reconstruction efficiency of
global tracks are not flat in azimuth and as a function of transverse
momentum, mostly due to missing or inefficient regions of the
ITS. These tracks are used for the \pt\ distribution measurements.
TPC-only tracks can be constrained to the primary vertex
(reconstructed also using the ITS information) to provide better
momentum resolution.

The data used for this analysis were collected in 2010, during the
first Pb--Pb run at the LHC, at a center-of-mass energy per nucleon
\snn~=~2.76~TeV.  The hadronic interaction rate was of the order of
100~Hz, low enough to avoid any space charge distortion effects in the
TPC~\cite{TpcTdr}.  
The trigger was provided by the \VZERO\ detector~\cite{Abbas:2013taa},
a pair of forward scintillator hodoscopes placed on either side of the
interaction region, covering the pseudorapidity regions
$2.8 < \eta <5.1$ (\VZEROA) and $-3.7 < \eta < -1.7$ (\VZEROC).
Events were requested to have a signal in both sides of the \VZERO,
selecting roughly 0--90\% most central
collisions~\cite{Abelev:2013qoq}.  The \VZERO\ measures a signal whose
average amplitude is proportional to the multiplicity of charged
particles. The \VZERO\ acceptance times detection efficiency is
approximately 90\% and flat as a function of the particle \pt, with
only a small reduction to about 85\% for $\pt < 300$~MeV/$c$. Events
are further selected offline using the timing information from the
\VZERO\ and from a set of two forward Zero Degree Calorimeters (ZDCs),
in order to reject contamination from beam-induced backgrounds
(see~\cite{Abelev:2013qoq,Aamodt:2010pb,Aamodt:2010cz} for a detailed
discussion).  After all selections, the event sample used in the
analysis consists of about 16 million events.

\section{Analysis technique}
\label{sec:analysis-technique}

\subsection{Centrality and the event shape selection}
\label{sec:event-selection}

The events which pass the basic selection described in
\Sect{sec:alice-detector-data} are divided in centrality classes
based on the signal amplitude (proportional to the charged particle
  multiplicity) measured in the \VZERO\ detector, as described in
~\cite{Abelev:2013qoq}. 
Events in each centrality class are further subdivided into groups
with different average elliptic event shapes based on the
magnitude of the second order reduced flow vector \qtwo~\cite{Voloshin:2008dg} given as
\begin{equation}
  \label{eq:small-q}
  \qtwo = \frac{\left| \pmb{Q}_{2} \right|}{\sqrt{M}},
\end{equation}
where $M$ is the multiplicity and
$\left| \pmb{Q}_{2} \right| = \sqrt{Q_{2,x}^2 + Q_{2,y}^2}$ is the
magnitude of the second order flow vector.

In this paper, the flow vector $\pmb{Q}_{2}$ is calculated using the
TPC or \VZERO\ detectors. In the TPC,
tracks in the range $0.2 < \pt < 20~\gevc$ and $|\eta|<0.4$
(to avoid an overlap with the $\eta$ region used for the \vtwo\ and
\pt\ distribution measurements) are used to measure
\begin{equation}
  \label{eq:big-q}
  Q_{2,x} = \sum_{i=1}^{M} \cos{2\varphi_i} ~ , 
~ Q_{2,y} = \sum_{i=1}^{M}  \sin{2\varphi_i},
\end{equation}
where $\varphi_i$ is the azimuthal angle of the $i$-th particle and
$M$ is the number of tracks in an event.

In the forward rapidity region the \VZERO\ is used. This
detector segmented into four rings, each consisting of 8 azimuthal
sectors, the flow vector is hence calculated as
\begin{equation}
  \label{eq:big-q-v0}
  Q_{2,x} = \sum_{i=1}^{32} w_i \cos{2\varphi_i} ~ , 
~ Q_{2,y} = \sum_{i=1}^{32} w_i \sin{2\varphi_i} ~ , ~ M = \sum_{i=1}^{32} w_i,
\end{equation}
where the sum runs over all 32 channels, $\varphi_i$ is the angle of
the center of the sector containing channel $i$, $w_i$ is the
amplitude measured in channel $i$ and $M$ is in this case the sum of
the amplitudes measured in each channel.

The discriminating power of \qtwo\ depends on the magnitude of
elliptic flow as well as on the track multiplicity used in the
\qtwo\ calculation and on the performance of the detector, including
the angular resolution or the linearity of the response to the charged
particle multiplicity. The good resolution of the \TPC\ and the large
multiplicity at midrapidity are used to maximize the selectivity on
\qtwo. However, the ALICE central barrel acceptance enables only
limited separation in pseudorapidity between the region used to
calculate \qtwo\ and the region used to calculate the observables
($|\Deta| = 0.1$). This separation is introduced in order to suppress
unwanted non-flow correlations, which typicaly involve only a few
particles and are in general short-range. In order to further assess
the contribution of non-flow correlations, the flow vector is also
calculated using the \VZERO\ detectors. This leads to a separation of
more than one unit in pseudorapidity between the two regions.

\begin{figure}[t!]
  \centering
  \includegraphics[width=0.9\textwidth]{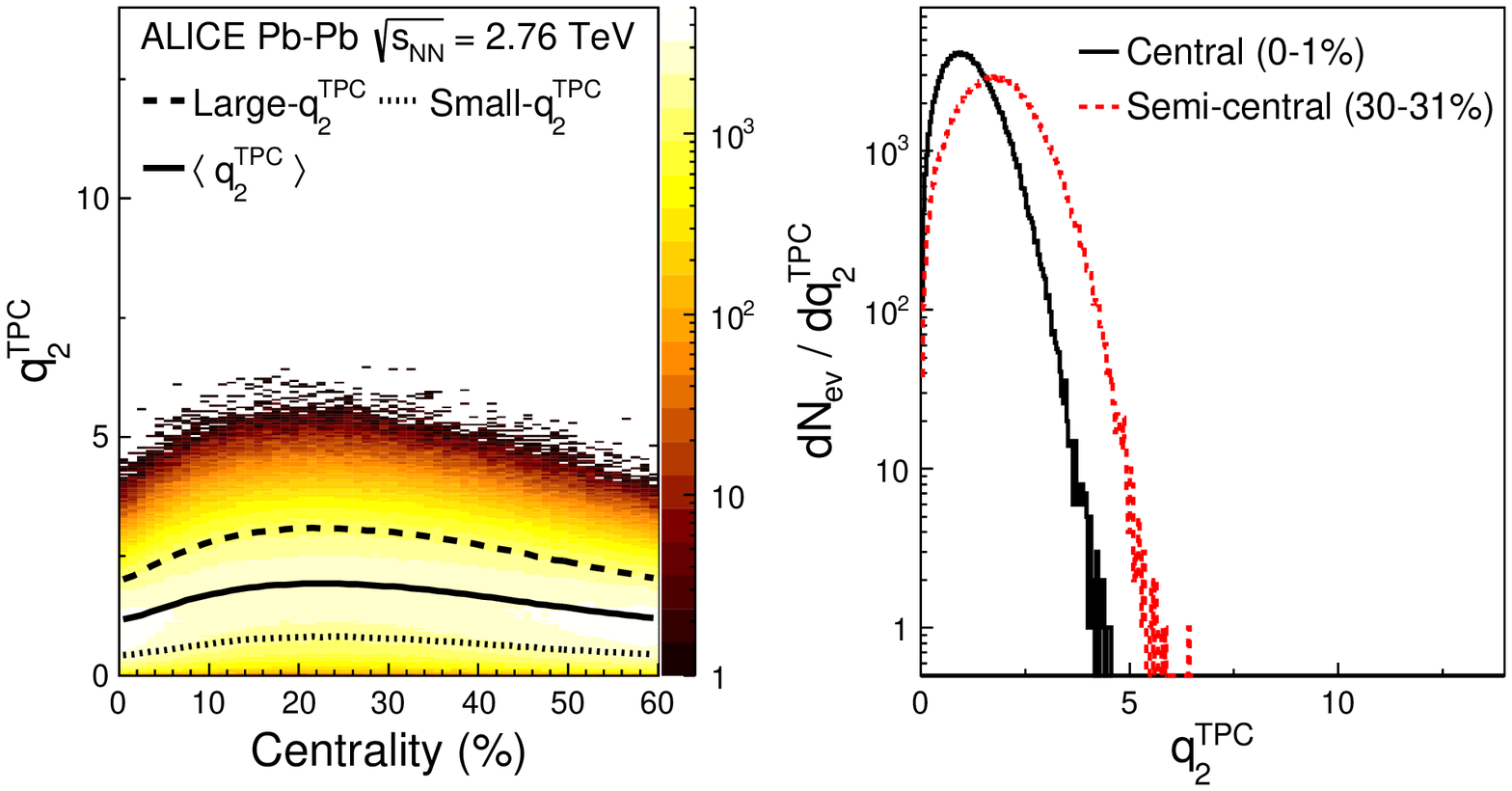}
  \includegraphics[width=0.9\textwidth]{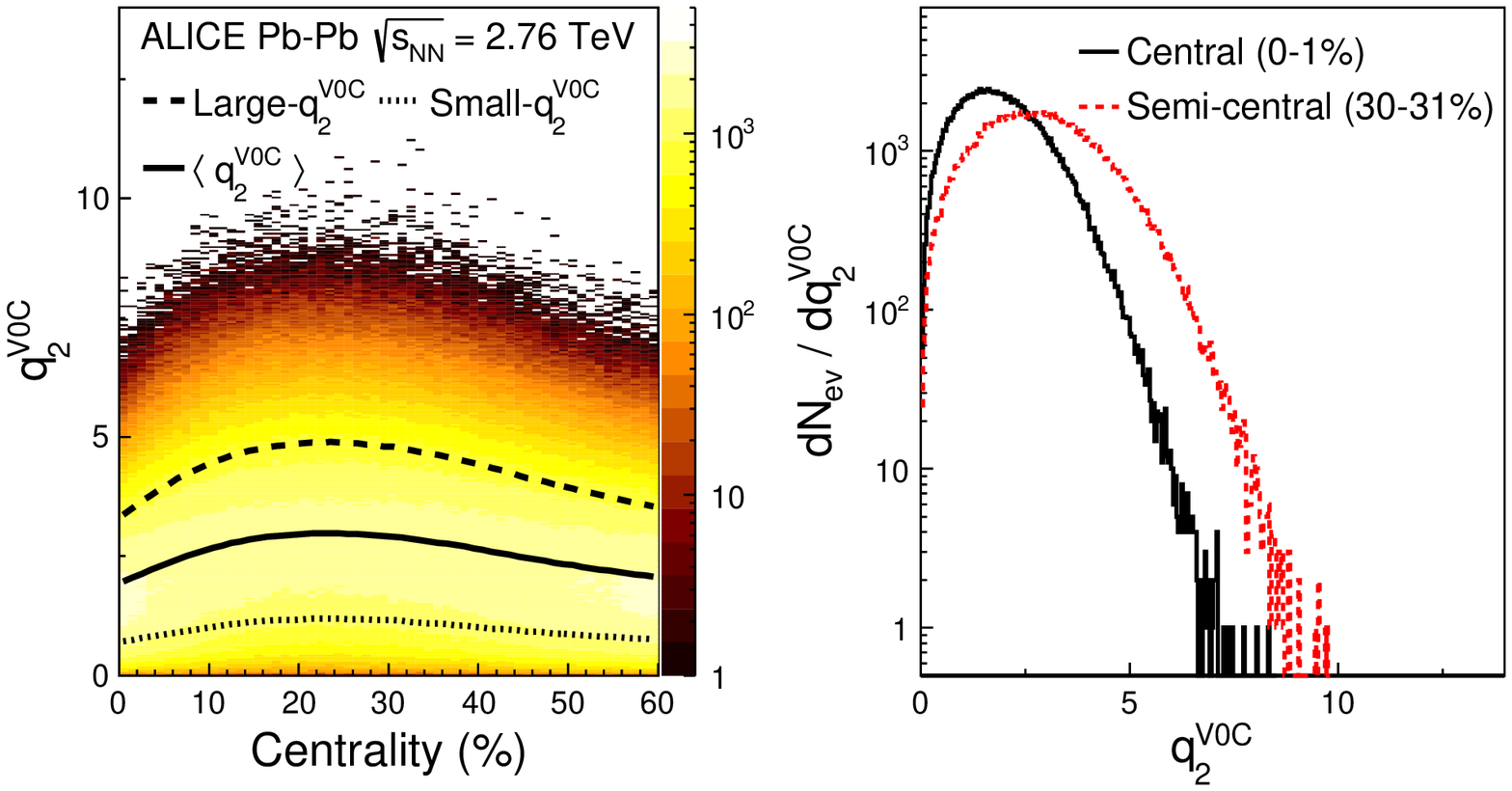}
  \caption{(Color online) Distributions of \qttpc\ (top) and \qtvc\ (bottom) as a
    function of centrality (left) and projections for two centrality
    classes, 0--1\% and 30--31\% (right). In each of the left panels
    the solid curve shows the average \qtwo\ as a
    function of centrality, while the dashed and the dotted curves
    indicate the top 10\% and the bottom 10\%, respectively.}
\label{fig:q-vector}
\end{figure}

In absence of correlations, the average length of $\pmb{Q}_{2}$ grows
as $\sqrt{M}$~\cite{Voloshin:2008dg}: \qtwo\ is introduced to remove
this trivial part of the multiplicity dependence.  In case of non-zero
correlations (due to either collective flow or non-flow correlations),
\qtwo\ depends on multiplicity and on the strength of the flow
as~\cite{Adler:2002pu,Voloshin:2008dg}
\begin{equation}
  \label{eq:q2-vs-mult}
  \left<\qtwo^2\right> \simeq 1 + \left<(M-1)\right>  
\left<\left( \vtwo^2+ \delta_2\right)\right>,
\end{equation}
where the parameter $\delta_2$ accounts for non-flow correlations, and
the angular brackets denote the average over all events.
                                
In the case when the multiplicity is measured via the signal amplitude
in the \VZERO\ detector, the first term in Eq.~\ref{eq:q2-vs-mult}
(unity) has to be substituted by $\langle e_i^2 \rangle/\langle
e_i\rangle^2$, where $e_i$ is the energy deposition of a single
particle $i$. The fluctuations in $e_i$ lead to an increase in the
flow vector length and reduce the corresponding event plane
resolution.

The \qtwo\ distribution measured with the \TPC\ (\qttpc) and \VZEROC\
(\qtvc) is shown in \Fig{fig:q-vector} as a function of centrality,
and in two narrow centrality classes, 0--1\% and 30--31\%. As can
be seen, \qtwo\ reaches values twice as large as
the mean value, as expected in case of large initial state
fluctuations~\cite{Schukraft:2012ah}. The \qtvc\ is larger than
\qttpc, as the former is measured in a larger pseudorapidity window
(integrating a larger multiplicity) and is sensitive to the
fluctuations in $e_i$. Note also that the selectivity (discrimination
power) of the two selection cuts is in principle different, due to the
different detector resolution, and, in the case of \VZEROC, smaller
\vtwo\ value at forward $\eta$, fluctuations in $e_i$ and large
contribution of secondary particles. 

\begin{figure}[t!]
  \centering
  \includegraphics[width=0.9\textwidth]{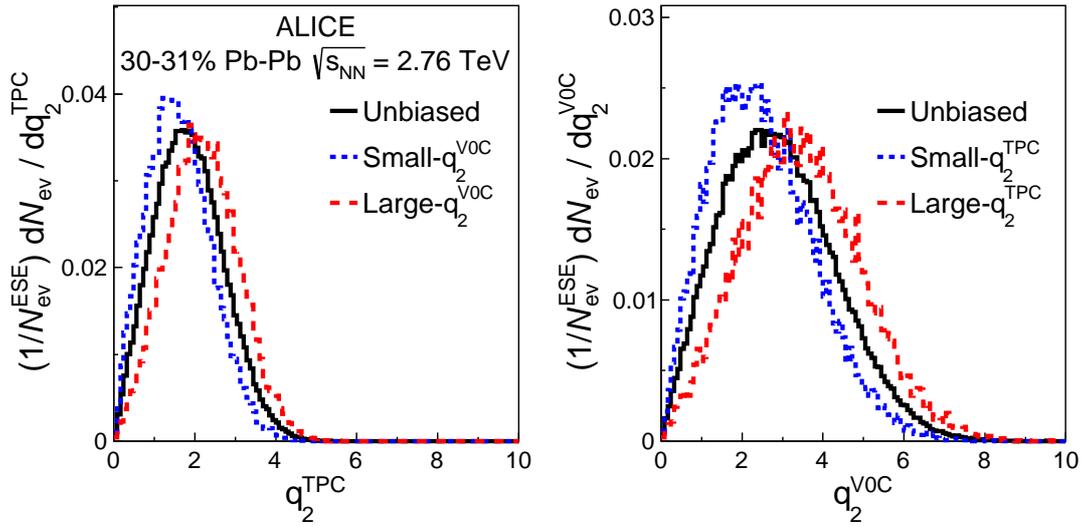}
  \caption{(Color online) Effect of the \qtvc\ (\qttpc) event shape selection on the
    \qttpc\ (\qtvc) distributions for events in the 30-31\% centrality
    class.}
\label{fig:q-vector-corr}
\end{figure}

In the present analysis, the effect of the event shape engineering on
\vtwo\ and \pt\ distributions is studied.  The average flow and
particle spectra are measured in the pseudorapidity range
$0.5<|\eta|<0.8$ in order to avoid overlap with the region used to
calculate \qttpc. The \VZEROC\ selection is used to estimate the
contribution of non-flow correlations to the event-shape selection,
since it provides a large $\eta$ gap.  As a further crosscheck, the
analysis was also repeated using the \VZEROA\ detector. The results
obtained with \VZEROA\ and \VZEROC\ show a qualitative agreement with
a better selectivity when the \VZEROC\ is used (mostly due to the
larger multiplicity in the acceptance of this detector and to the
$\eta$ dependence of the elliptic flow). We therefore report the
results for events selected using \qttpc\ and \qtvc\ in this paper.

Due to the limited statistics, the analysis has to be performed in
relatively wide centrality classes ($\sim$~10\%). The length of \qtwo\
changes within such large centrality intervals (\Fig{fig:q-vector}),
and a cut at a fixed value of \qtwo\ would introduce a dependence on
the multiplicity that would obscure the effect of the event-shape
selection. The \qtwo\ selection is therefore evaluated in narrow
(1\%-wide) centrality classes.  The results presented in the next
sections are obtained in two event-shape classes, corresponding to the
10\% of the events having the top (bottom) value of the \qtwo\
(estimated in the narrow centrality
classes). 
In the following, we refer to these two classes as ``\hq''
(90-100\%) and ``\sq'' (0-10\%) or, generically, as ESE-selected
events.  Conversely, we refer to the totality of data within a given
centrality class as the ``unbiased'' sample.

The correlation between \qttpc\ and \qtvc\ is illustrated for events
in the 30-31\% centrality class in \Fig{fig:q-vector-corr}. The left
(right) panel shows the distribution of \qtwo\ measured with the
\TPC\ (\VZEROC) for all events and for events in the \hq\ and
\sq\ classes, selected with the \VZEROC\ (\TPC).  
The average \qtwo\ changes by about 18\% and 14\% in the
\hq\ and \sq\ samples respectively. In order to control the effect
of fluctuations in a given detector the detailed comparison of the
results obtained with \qttpc\ and \qtvc\ is crucial, as discussed in
detail below.
In order to disentangle the effect of the $\eta$ gap and of the
\qtwo\ cut, the selection on \qttpc\ is also adjusted such that the
average flow measured at mid-rapidity is similar to the one in the
\hq\ sample (\Sect{sec:results}).

The ESE becomes less selective in peripheral events regardless of the
detector used to compute \qtwo, due to the low multiplicity. This
limits the present analysis to the 60\% most central events.

Space charge distortion effects in the TPC, which accumulate over many
events could, in principle, bias the \qtwo\ selection.  In order to
check for this and other possible instrumental effects it was verified
that the results are not sensitive to the instantaneous luminosity.

\subsection{Elliptic flow measurement}
\label{sec:flow-measurement}

The elliptic flow, \vtwo, is measured in the pseudorapidity range $0.5
< \left|\eta\right| < 0.8$ using the Scalar Product (SP)
method~\cite{Voloshin:2008dg}, according to:
\begin{equation}
\label{eq:v2SP}
  v_{2}\{\mathrm{SP}\} = \frac{\langle \
  	{\pmb{u}}_{2,k} \pmb{Q}^{*}_{2} / M \rangle }
{\sqrt{\pmb{Q}^{A}_{2}\pmb{Q}^{B*}_{2} / M^{A} M^{B}}}
\end{equation}
where $\pmb{u}_{2,k}$=exp(i2$\varphi_{k}$) is the particle's unit flow
vector, $\varphi_{k}$ is the azimuthal angle of the $k$-th particle of
interest, $\pmb{Q}_{2}$ is the flow vector and $M$ is the
multiplicity. The full event is divided in two independent sub-events,
labeled as $A$ and $B$, covering two different pseudorapidity ranges,
$0.5 < \eta < 0.8$ and $-0.8 < \eta < -0.5$. The particle's unit flow
vector $\pmb{u}_{2,k}$ is evaluated in the sub-event $A$ while the
flow vector $\pmb{Q}_{2}$ and the multiplcity $M$ in the sub-event $B$
and vice-versa, ensuring a pseudorapidity gap of $|\Deta| > 1$ between
the particle of interest and the reference charged particles, which
suppresses the non-flow contribution in the calculation of \vtsp. A
flat acceptance in azimuth is achieved in this analysis selecting
TPC-only tracks, constrained to the primary vertex. Tracks are
required to have at least 70 clusters and a $\langle\chi^{2}\rangle
\leq 4$ per TPC cluster (two degrees of freedom). Tracks with a
transverse distance of closest approach to the vertex (computed before
constraining tracks to the primary vertex) $\dcaxy~>~2.4~\rm{cm}$ or a
longitudinal distance of closest approach $\dcaz~>~3.2~\rm{cm}$ are
rejected to reduce the contamination from secondary tracks.  The
effect of secondary particles is corrected applying the same analysis
procedure to Monte Carlo events, simulated with the AMPT event
generator~\cite{Lin:2004en} and propagated through a
GEANT3~\cite{Geant:1994zzo} model of the detector. The \vtsp\ computed
using reconstructed tracks is then compared with the one computed with
generated primary particles, and the difference ($<5\%$) is used as a
correction factor.\warn{shall we mention pt dependence?}

The uncertainty on the tracking efficiency was assessed with different
track samples and selections: using a set of hybrid tracks, built from
a combination of global and TPC only tracks to obtain a uniform azimuthal acceptance~\cite{Abelev:2013kqa},
using TPC only tracks not constrained to the primary vertex, varying
the minimum number of TPC clusters required in the analysis from 70 to
50 (\emph{Track reconstruction} in Tab.~\ref{tab:v2_syst_tab} and
\ref{tab:v2ese_syst_tab}) and weighting each track by the inverse of
the (\pt-dependent) efficiency (\emph{Tracking efficiency}).

The procedure used to estimate the centrality percentiles leads to a
$\sim$1\% uncertainty in the definition of the centrality
classes~\cite{Abelev:2013qoq}.  In order to propagate this uncertainty
to the results presented in this paper, the measurement is repeated
displacing the centrality percentile by 1\%. For instance, the
analysis in the 30-40\% centrality class is repeated for the selection
30.3-40.4\% (\emph{Centrality resolution}).  Moreover, tracks
reconstructed at midrapidity (instead of the \VZERO\ signal) are used
as the centrality estimator (\emph{Centrality estimator}).

The correction for the effect of secondary particles mentioned above
is strongly model dependent, therefore the difference between the
\vtwo\ estimated using generated AMPT particles and reconstructed
tracks was used to estimate the corresponding systematic uncertainty,
$\sim$ 3.5\% (0.7\%) at $\pt = 0.2\ (1.5)$~\gevc\ (\emph{Secondary
  particles}).

Moreover, the following systematic checks were considered: the
dependence on the magnetic field configuration was studied analyzing
separately samples of events collected with different polarities of
the magnetic field (\emph{Magnetic field}), analyzing positive and
negative particles separately (\emph{charge)} and analyzing samples of
tracks produced at different vertex positions: $-10 < \zvtx < 0 $ cm
and $0 < \zvtx < 10 $ cm (\emph{Vertex}). These effects are found to
be not significant.

\begin{table}
\center
\begin{tabular}{c c c c}
\hline                
\hline
Effect                 & \vtwo\ & \vtwo\ \hq\ & \vtwo\ \sq\ \\
\hline
\hline
Track reconstruction   & 3.1\% (0-20\%)  & 3.1\% (0-20\%)  & 3.1\% (0-20\%)    \\ 
                       & 2.7\% (20-60\%) & 2.7\% (20-60\%) & 2.7\% (20-60\%)   \\ 
                       & (\pt=0.2 \gevc)  & (\pt=0.2 \gevc)  & (\pt=0.2 \gevc) \\ \cline{2-4}
                       & 0.08\% (0-20\%)  & 0.08\% (0-20\%)  & 0.08\% (0-20\%)    \\ 
                       & 0.02\% (20-60\%) & 0.02\% (20-60\%) & 0.02\% (20-60\%)   \\ 
                       & (\pt=1.5 \gevc)  & (\pt=1.5 \gevc)  & (\pt=1.5 \gevc) \\ \cline{1-4}

Tracking efficiency    & 0.07\%     & 0.35\%   & 0.14 \% \\ \cline{1-4}
Centrality resolution  & 0.21\%     & 0.35\%   & 0.35\%    \\ \cline{1-4}
Centrality estimator   & 0.57\%     & 0.49\%   & 0.57\% \\ \cline{1-4}
Secondary particles    & 3.56\%     & 3.56\%   & 3.56\%  \\
& ($\pt~=~0.2~\gevc$) & ($\pt~=~0.2~\gevc$) & ($\pt~=~0.2~\gevc$) \\ \cline{2-4}
                       & 0.8\%      & 0.8\%    & 0.8\%   \\
                       & ($\pt~=~1.5~\gevc$) &  ($\pt~=~1.5~\gevc$) & ($\pt~=~1.5~\gevc$) \\ \cline{1-4}
       
Magnetic field         & NS         & NS       & NS       \\ \cline{1-4}
Charge                 & NS         & NS       & NS       \\ \cline{1-4}
Vertex                 & NS         & NS       & NS       \\ \cline{1-4}
\hline                      
\hline	
\end{tabular}
\caption{Summary of systematic errors on \vtsp\ measurement. NS = not statistically significant.}
\label{tab:v2_syst_tab}

\vspace{0.5cm}

\begin{tabular}{c c c c}
\hline                
\hline
Effect                 & \vtwo\ \hq/unbiased         & \vtwo\ \sq/unbiased \\
\hline
\hline
Track reconstruction   & 0.14\%              & 0.14\%     \\ \cline{1-3}
Tracking efficiency    & 0.35 \%             & 0.21\%      \\ \cline{1-3}
Centrality resolution  & 0.14\%              & 0.21\%     \\ \cline{1-3}
Centrality estimator   & 0.14\%              & 0.07\%      \\ \cline{1-3}
Secondary particles    & 0.07\%              & 0.35\%  \\ \cline{1-3}          
Magnetic Field         & NS                  & NS         \\ \cline{1-3}
Charge                 & NS                  & NS         \\ \cline{1-3}
Vertex                 & NS                  & NS         \\ \cline{1-3}
\hline                
\hline
\end{tabular}
\caption{Summary of systematic errors on the \vtsp\ ratios. NS = not statistically significant.}
\label{tab:v2ese_syst_tab}
\end{table}

The systematic uncertainties in the \vtwo\ measurements and in the
ratios of \vtwo\ in ESE-selected over unbiased events are summarized
in Tab.~\ref{tab:v2_syst_tab} and~\ref{tab:v2ese_syst_tab}. Only the
checks and variations which are found to be statistically significant
are considered in the systematic
uncertainties~\cite{Barlow:2002yb}. Whenever the \pt\ dependence of the
uncertainty is not negligible, values for characteristic \pt\ are
given in the tables.

\subsection{Transverse momentum distribution measurement}
\label{sec:spectra-measurement}

The measurement of the \pt\ distributions uses global tracks, which
provide good resolution on \dcaxy\ (\Sect{sec:alice-detector-data}),
and hence good separation of primary and secondary particles.  The
track selection requires at least 70 clusters in the TPC and at least
2 points in the ITS, out of which at least one must be in the first
two layers, to improve the \dcaxy\ resolution. A \pt-dependent cut on
the \dcaxy, corresponding to 7 times the experimental resolution on
\dcaxy, is applied to reduce the contamination from secondary
particles. Tracks with a $\chi^{2}$ per point larger than 36 in the
ITS and larger than 4 in the TPC are rejected. Finally, to further
reduce the contamination from fake tracks, a consistency cut between
the track parameters of TPC and global tracks was applied. For each
reconstructed TPC track, the $\chi^2$-difference between the track
parameters computed using only the TPC information constrained to the
vertex and the associated global track is required to be less than
36~\cite{Abelev:2012hxa}. Charged tracks are studied in the
pseudorapidity window $0.5 < \left|\eta\right| < 0.8$, to avoid an
overlap with the \qttpc\ calculation.

Particles are identified using the specific energy loss \dedx\ in the
TPC and their arrival time in the TOF\@. The technique is similar to
the one presented in~\cite{ABELEV:2013wsa}. A track is identified as
either a pion, a kaon or a proton based on the difference, in the
detector resolution units, from the expected energy loss and/or time
of flight $\nsigma^i_{\rm PID}$ (with $i$ being the particle identity
under study).  Below \pt~=~0.5~\gevc, only the TPC information is used
($\nsigma^i_{\rm PID}=\nsigma^i_{\rm TPC}$). For larger \pt, the TPC
and TOF information is combined using a geometrical mean:
$\nsigma^i_{\rm PID} = \sqrt{(\nsigma^i_{\rm TPC})^2+(\nsigma^i_{\rm
    TOF})^2}$.  Tracks are required to be within $3\sigma_{\rm PID}$
of the expected value to be identified as $\pi^{\pm}$, K$^\pm$ or p
(\pbar). In the region where the 3$\sigma_{\rm PID}$ identification
bands of two species overlap, the identity corresponding to the
smaller $\nsigma_{\rm PID}$ is assigned.  This technique gives a good
track-by-track identification in the following \pt\ ranges:
$0.2<\pt<4~\gevc$ for $\pi^\pm$, $0.3<\pt<3.2~\gevc$ for K$^\pm$,
$0.5<\pt<4~\gevc$ for p (\pbar). The misidentification of tracks is
below 4\% for pions, 25\% for kaons and 10\% for protons in those
ranges. Further discussion on the ALICE Particle Identification (PID)
performance can be found in~\cite{Abelev:2014ffa,Abelev:2013vea}. The
results for identified particles are provided in the pseudorapidity
range $0.5 < \left|\eta\right| < 0.8$. However, in the case of the
\qtvc\ selection the results were also studied at mid-rapidity
$\left|y\right| < 0.5$. Results for positive and negative particles
are consistent. In the following, ``pions'', ``kaons'' and
``protons'', as well as the symbols ``$\pi$'', ``K'' and ``p'', refer
to the sum of particles and antiparticles.

\begin{table}
\center
\begin{tabular}{ c c c c c}
\hline
\hline
Effect                 & $\rm N_{ch}$ & $\pi^{\pm}$ & $K^{\pm}$ & p and $\bar{\rm p}$ \\ \cline{1-5}
\hline
\hline

Track reconstruction   & $<$ 0.035\%   & 0.07\%      & 0.07\%     & 0.07\%         \\ \cline{1-5}

Tracking efficiency    & 0.21\%       & 0.21\%      & 0.21\%     & 0.21\%         \\ \cline{1-5}

Centrality resolution  & 0.07\% ($\pt> 1.5~\gevc$)  & 0.07\% ($\pt> 1.5~\gevc$)  & 0.14\% & 0.14\% \\ \cline{1-5}

Centrality estimator   & 0.35\%  & 0.35\%  & 0.35\% & 0.35\% \\ \cline{1-5}

PID                    & -     & 0.07\% ($\pt> 1.5~\gevc$)  & 0.07\%  & 0.07\%    \\ \cline{1-5}

Secondary particles    & $<$ 0.035\%     & $<$ 0.035\% & $<$ 0.035\% & 0.07\%    \\ \cline{1-5}
Normalization          & 1.1\%  & 1.1\%  & 1.1\%  & 1.1\% \\ \cline{1-5}
Magnetic field         & NS     & NS           & NS         & NS                  \\ \cline{1-5}

Charge                 & $<$ 0.035\%   & $<$ 0.035\%  & $<$ 0.035\%  & $<$ 0.035\%    \\ \cline{1-5}

Vertex                 & 0.07\%       & 0.07\%      & 0.07\%      & 0.07\%        \\ \cline{1-5}
\hline
\hline
\end{tabular}
\caption{Summary of systematic errors for the ratio of \pt\ distributions between large-\qtwo\ and unbiased events. NS = not statistically significant.}
\label{tab:ese_syst_lq}

\vspace{0.5cm}

\begin{tabular}{ c c c c c}
\hline
\hline
Effect                 & $\rm N_{ch}$ & $\pi^{\pm}$ & $K^{\pm}$ & p and $\bar{\rm p}$ \\ \cline{1-5}
\hline
\hline
Track reconstruction   & $<$ 0.035\%   & 0.07\%      & 0.07\%     & 0.07\%         \\ \cline{1-5}

Tracking efficiency             & 0.28\%  & 0.28\%  & 0.28\%  & 0.28\% \\ \cline{1-5}

Centrality resolution  & 0.07\% ($\pt> 1.5~\gevc$)  & 0.07\% ($\pt> 1.5~\gevc$)  & 0.14\% & 0.14\% \\ \cline{1-5}

Centrality estimator   & 0.35\%  & 0.35\%  & 0.35\% & 0.35\% \\ \cline{1-5}

PID                    & -     & 0.07\% ($\pt> 1.5~\gevc$)  & 0.07\%  & 0.07\%    \\ \cline{1-5}
Secondary particles    & $<$ 0.035\%     & $<$ 0.035\% & $<$ 0.035\% & 0.07\%    \\ \cline{1-5}

Normalization          & 0.6\%  & 0.6\%  & 0.6\%  & 0.6\% \\ \cline{1-5}
Magnetic field         & NS     & NS           & NS         & NS                  \\ \cline{1-5}

Charge                 & $<$ 0.035\%   & $<$ 0.035\%  & $<$ 0.035\%  & $<$ 0.035\%    \\ \cline{1-5}

Vertex                 & 0.07\%       & 0.07\%      & 0.07\%      & 0.07\%        \\ \cline{1-5}

\hline
\hline
\end{tabular}
\caption{Summary of systematic errors for the ratio of \pt\ distributions between small-\qtwo\ and unbiased events.  NS = not statistically significant.}

\label{tab:spectra_syst_tab}
\end{table}

The results for the spectra in ESE-selected events are presented in
terms of ratios between the distributions measured in the \hq\ (\sq)
samples and the unbiased sample. The unbiased spectra have already
been reported in~\cite{Abelev:2012hxa,Abelev:2013vea}. Most of the
corrections (and uncertainties) cancel out in these ratios, allowing
for a precise determination of the effect due to the event-shape
selection, as discussed in detail below. The uncertainties can mostly
arise due to effects that depend on the local track density, which are
found to be small~\cite{Abelev:2014pua}.
  
The systematic uncertainties are summarized in
Tab.~\ref{tab:ese_syst_lq} and Tab.~\ref{tab:spectra_syst_tab}. As
mentioned before, only the checks and variations which are found to be
statistically significant are considered in the systematic
uncertainties~\cite{Barlow:2002yb}.

The systematic uncertainty related to the tracking is estimated
varying the track selection cuts. Instead of the standard TPC cluster
cut, at least 120 (out of 159) pad-rows hits in the TPC and a fraction
of shared clusters in the TPC $< 0.4$, are required (\emph{Track
  reconstruction} in Tab.~\ref{tab:ese_syst_lq} and
\ref{tab:spectra_syst_tab}).
  
The possible effect of a track-density-dependent efficiency (which
would influence in a different way events with the large$-$ and
\sq\ selection) is investigated using simulations based on the AMPT
event generator~\cite{Lin:2004en} and a parametric event generator tuned
to reproduce the ALICE spectra and
\vtwo\ measurements~\cite{Abelev:2014pua}. This effect leads to an
uncorrelated systematic error of about 0.2\% and a normalization error
of 0.4\% (\emph{Tracking efficiency}).

The uncertainty on the centrality is estimated varying the definitions
of centrality classes by 1\% and using tracks as the centrality
estimator. These checks lead to an uncorrelated uncertainty of about
0.1\% and 0.35\%, respectively and a normalization uncertainty below
1\% in the ratios of spectra (\textit{Centrality resolution} and
\textit{Centrality estimator}).

The systematic effect related to the particle identification is
studied performing several variations to the PID approach described
above.  The $\nsigma_{\rm PID}$ cut is varied between 2 and 4.
Alternatively, if a track is consistent with more than one particle
assignment within the $\nsigma_{\rm PID}$ cut, double counting is
allowed. As compared to the standard strategy where only the identity
closest to the measured $\nsigma_{\rm PID}$ is selected, this approach
leads to a slightly larger contamination from misidentified tracks,
but also to a larger efficiency. Finally, an exclusive $\nsigma_{\rm
  PID}$ strategy was used, which drastically reduces
misidentification: a particle is only accepted if it is compatible
with only one mass hypothesis at 3$\sigma_{\rm PID}$. As a further
cross-check, a Bayesian approach~\cite{Abelev:2014ffa} was also
considered. This method allows for better control of contamination at
high \pt. Overall, the uncertainty related to the particle
identification strategy is less than 0.1\% (\emph{PID}).

The effect of secondary particles depends on the \pt\ distribution of
weakly-decaying primary particles, and could be different for the
\hqnoq\ and \sq\ samples.  This effect is estimated to be at most
$\sim 0.1\%$ for protons with the \TPC\ ESE selection and negligible
in all other cases (\emph{Secondary particles}).

Possible effects related to the magnetic field and to the charge state
are addressed studying separately events collected with different
magnet polarities (\emph{Magnetic field}) and different charges
(\emph{Charge}), as in the case of the \vtsp\ measurement.  Particles
produced at different longitudinal position cross a different portion
of the detector, with different reconstruction efficiency. The samples
of events produced with a negative ($-10 < \zvtx < 0 $ cm) and
positive ($0 < \zvtx < 10$ cm) longitudinal vertex coordinate with
respect to the nominal interaction point were studied separately
(\emph{Vertex}).

\section{Results}
\label{sec:results}

\subsection{Charged particle elliptic flow}
\label{sec:flow-results}

\begin{figure}[p]
  \centering
  \includegraphics[width=\textwidth]{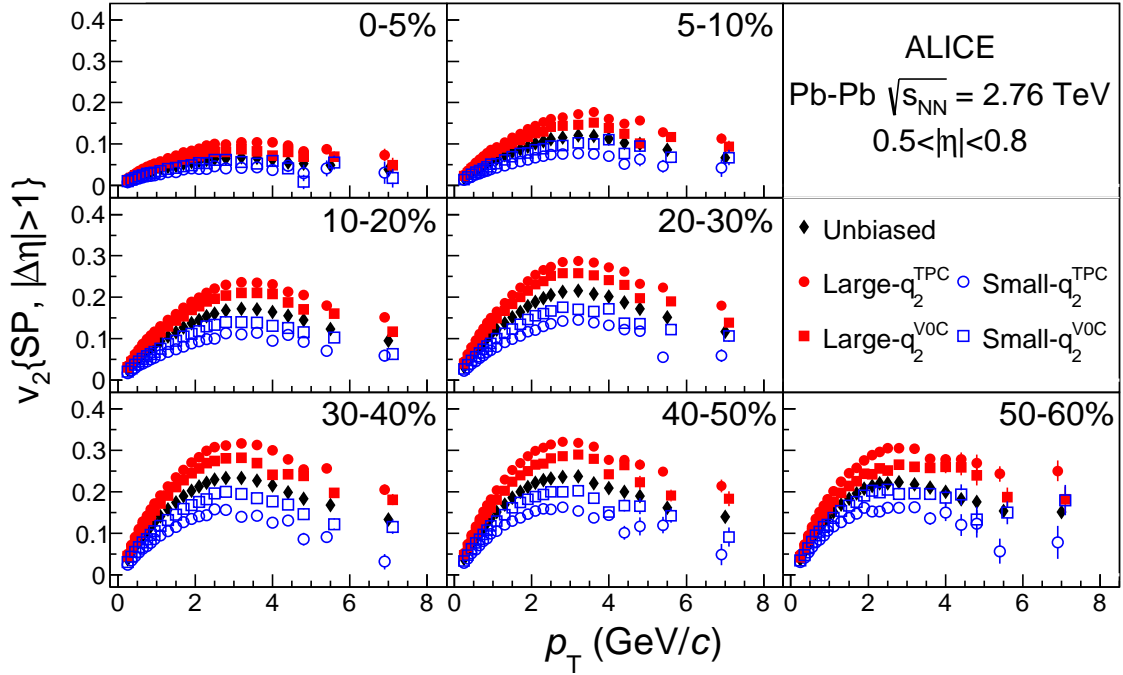}
  \caption{(Color online) Measurement of \vtsp\ as a function of \pt\ in different centrality
    classes for the unbiased, the \hq\ and the \sq\ samples. Only
    statistical uncertainties are plotted (systematic uncertainties are
    smaller than the markers).}
  \label{fig:v2ese}
\end{figure}

\begin{figure}[p]
  \centering
  \includegraphics[width=\textwidth]{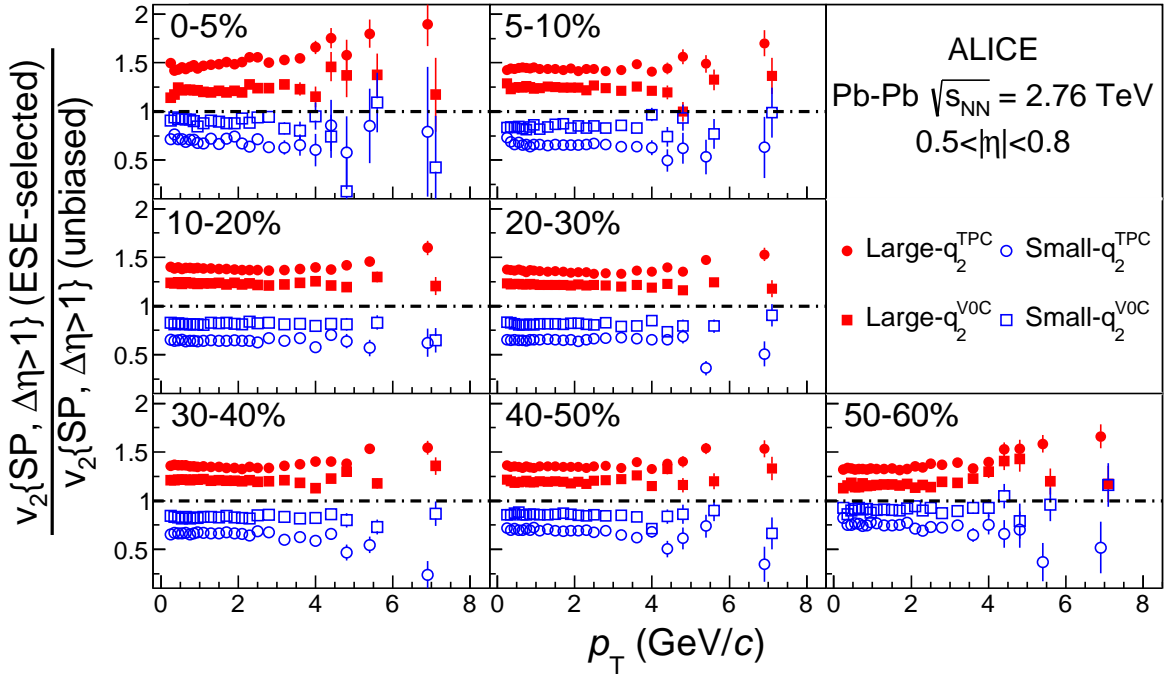}
  \caption{(Color online) Ratio of \vtsp\ in the \hq\ and \sq\ samples to unbiased
    sample. Only statistical uncertainties are plotted (systematic
    uncertainties are smaller than the markers).}
  \label{fig:v2ese-ratio}
\end{figure}

The event-shape selection is studied in \Fig{fig:v2ese}, where the
\vtsp\ as a function of \pt\ is reported for the unbiased and
ESE-selected samples, with both the \qttpc\ ($\left|\eta\right|<0.4$)
and \qtvc\ ($-3.7 < \eta < -1.7$) selections in different centrality
classes. \Figure{fig:v2ese-ratio} shows the ratio between the \vtwo\
measured with the \hq\ (\sq) selection and the unbiased
sample. Selecting the 10\% highest (lowest) \qttpc\ samples leads to a
change of 30-50\% in the \vtsp\ measured, depending on centrality. The
change is smaller ($\sim$ 10-25\%) in the case of \qtvc-based
selection, as compared to the \qttpc\ case.  As already indirectly
inferred from the difference between $\mathrm{2^{nd}}$ and
$\mathrm{4^{th}}$ order flow cumulants $v_{2}\{2\}$ and $v_{2}\{4\}$
in~\cite{Abelev:2012di}, the elliptic flow response of the system to
geometry fluctuations is almost independent of \pt.  For all
centralities, the change observed in Fig.~\ref{fig:v2ese-ratio}
depends indeed weakly on \pt, up to at least 4-5~\gevc.  This
indicates that a cut on \qtwo\ selects a global property of the event,
likely related to the initial shape in the overlap region. The only
exception to the previous observation is the 0-5\% centrality class,
where for the \qttpc\ selection an increasing trend with \pt\ is
observed. In this centrality class the mean value of \vtwo\ is small,
due to the almost isotropic shape in the initial state.  Moreover,
relative flow fluctuations are large in central collisions, with a
\pt\ dependence similar to the one shown in Fig.~\ref{fig:v2ese-ratio}
\cite{Abelev:2012di}. The analysis of the \pt\ spectra presented
in \Section{sec:spectra-results} gives additional insight into the
trend observed in \Figure{fig:v2ese-ratio}.

\begin{figure}[t!]
  \centering
  \includegraphics[width=0.8\textwidth]{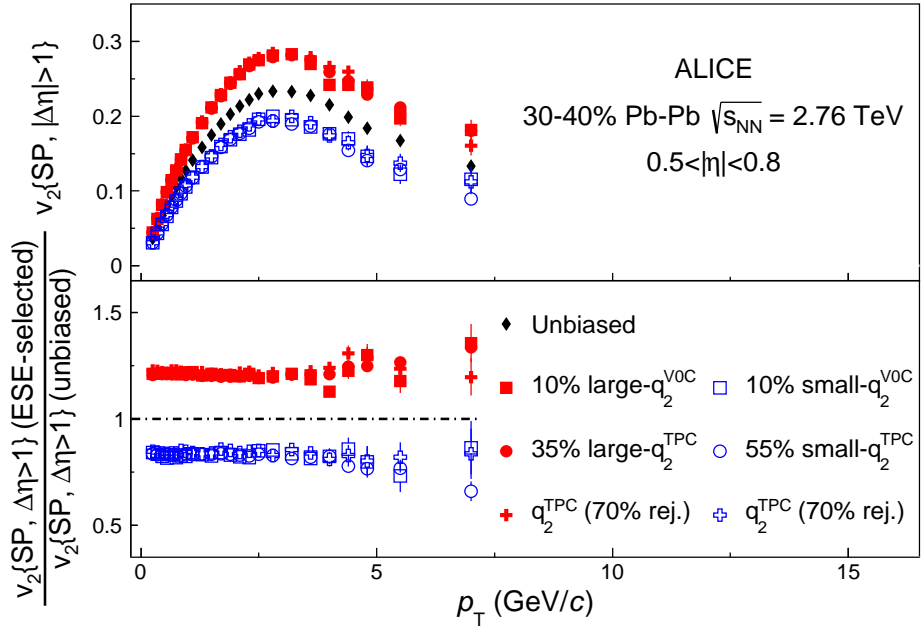}
  \caption{(Color online) Comparison between the effect of the event-shape selection
    obtained with the standard \VZEROC\ and with the tuned
    \TPC\ selections (see text for details), in the centrality class
    30-40\%. Top: \vtsp, bottom: ratios to the unbiased sample. Only
    statistical uncertainties are plotted (systematic uncertainties
    are smaller than the markers).}
  \label{fig:v2ese-relaxed}
\end{figure}

For $\pt \gtrsim 4-5~\gevc$, the ratio ESE-selected/unbiased
\vtsp\ increases for the \hq\ selection. This trend is more pronounced
for the \qttpc\ selection and for the most central and the most
peripheral classes. A fit with a constant over the full \pt\ range
yields $\chi^2$ per degree of freedom values in the range 2-6
(depending on centrality) for the \qttpc\ selection and $<2$ for the
\qtvc\ selection.  Fitting the ranges $\pt<5~\gevc$ and $\pt>5~\gevc$
with two different constants indicates an increase for the
\hq\ selection of order 5\% and 10\% for the \qtvc\ and
\qttpc\ selections, respectively. This difference could be due to a
small non-flow-induced bias. At high \pt\ the \vtwo\ is believed to be
determined by the path-length dependence of parton energy
loss~\cite{Abelev:2012di}.

The difference between the \qttpc\ and \qtvc\ can be due to the
different selectivity (see \Sect{sec:event-selection}), but also to a
different contribution of non-flow correlations between the \qtwo\ and
the \vtwo\ measurements. Replacing the \qttpc\ selection with the
\qtvc\ one changes both non-flow and selectivity at the same time. To
disentangle these two contributions, the selectivity of the \qttpc\
selection was artificially reduced.  This is achieved either relaxing
the selection itself or rejecting a random fraction of tracks for the
computation of \qttpc, while still selecting 10\% of the events. It is
found that selecting the class 65-100\% for the \hq\ sample (0-55\%
for the \sq\ sample) with \qttpc, or alternatively rejecting 70\% of
the TPC tracks, leads to an average variation of the \vtsp\ in the
range $0.2<\pt<4~\gevc$ comparable to the one obtained with the
standard 10\% \qtvc\ selection.  The results are shown in
Fig.~\ref{fig:v2ese-relaxed} for the centrality class 30--40\%. Not
only is it possible to find a cut which leads to the same average
variation in \vtsp, but the \pt\ dependence is very similar in both
cases.  Rejecting randomly 70\% of the tracks changes the selectivity
of \qttpc\ without affecting non-flow correlations between the \qttpc\
selection and \vtsp\ measurement (as the $\eta$ gap is not
varied). Also in this case, it is found that the effect of the \qtwo\
selection does not depend on \pt. A similar result, with the same
value of the relaxed cut or fraction of rejected tracks, is found for
the centrality interval 10--50\%. Moreover, as it will be discussed in
the next section, the same relaxed selections lead to the
same effect on the \pt\ distributions.

\begin{figure}[t!]
  \centering
  \includegraphics[width=0.8\textwidth]{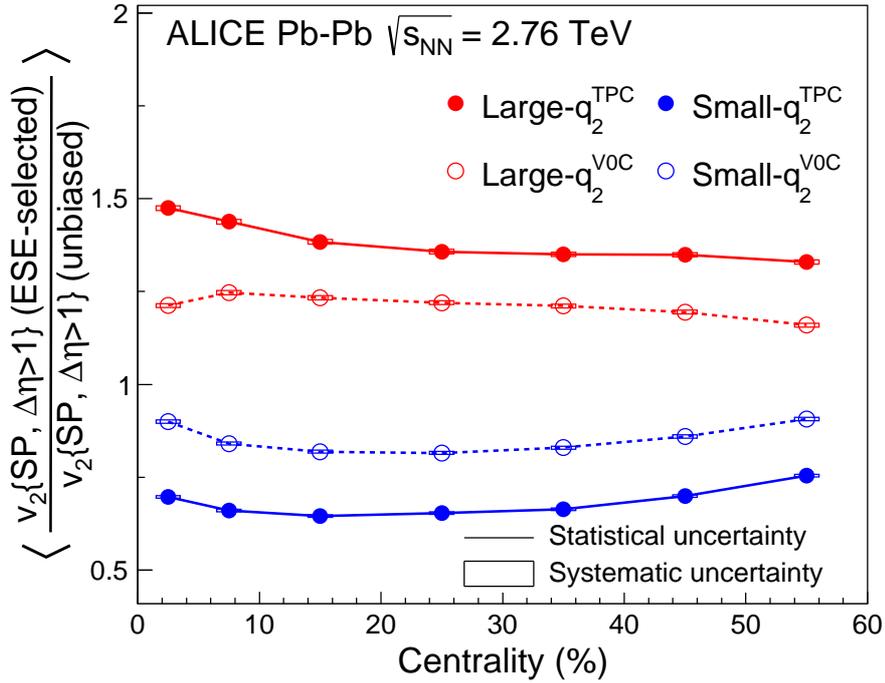}
  \caption{(Color online) Centrality dependence of the average \vtsp\ variation in
    the \hq\ and \sq\ samples.}
  \label{fig:v2ese-vs-centr}
\end{figure} 

\begin{figure}[t!]
  \centering
  \includegraphics[width=0.48\textwidth]{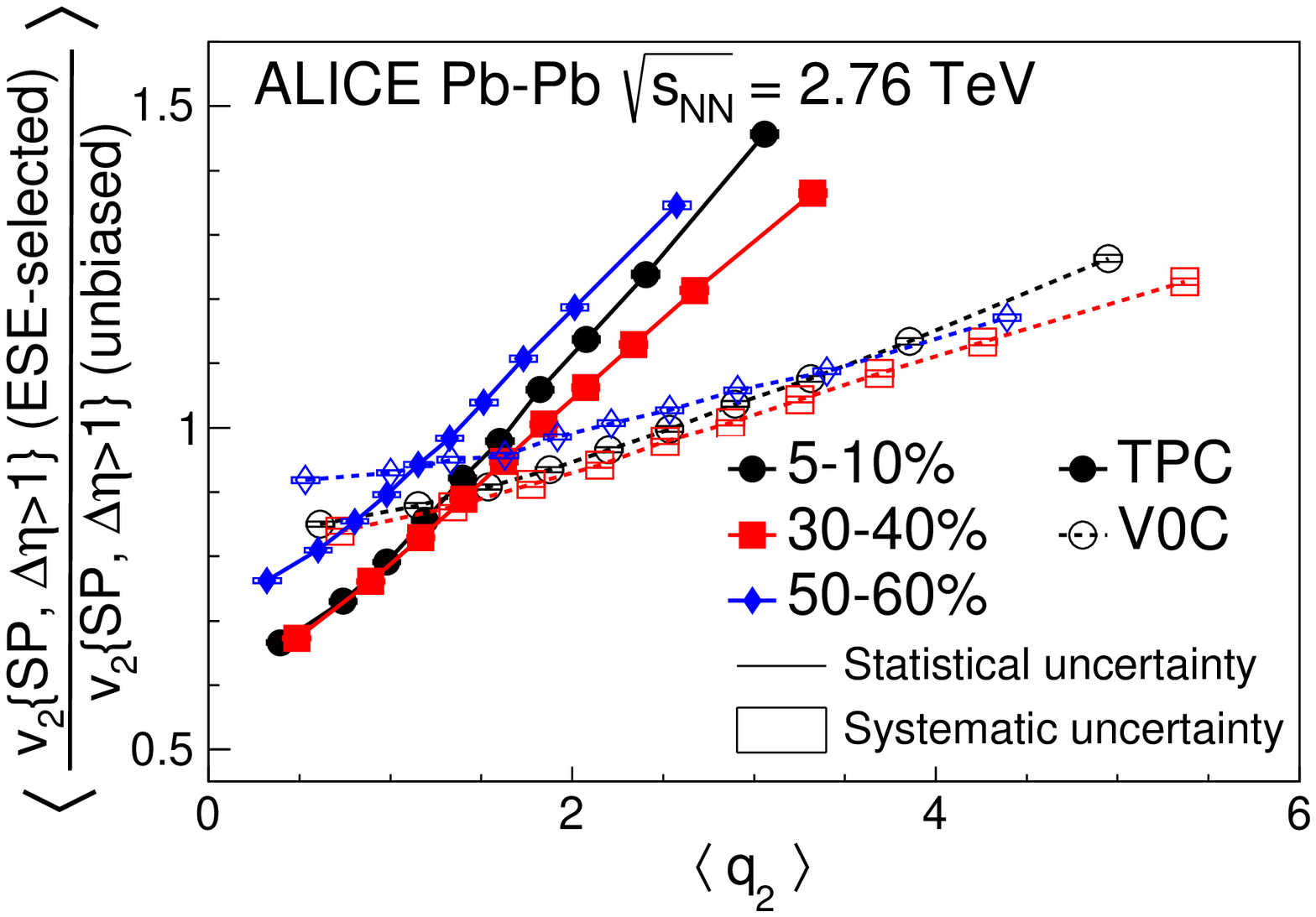}
  \includegraphics[width=0.48\textwidth]{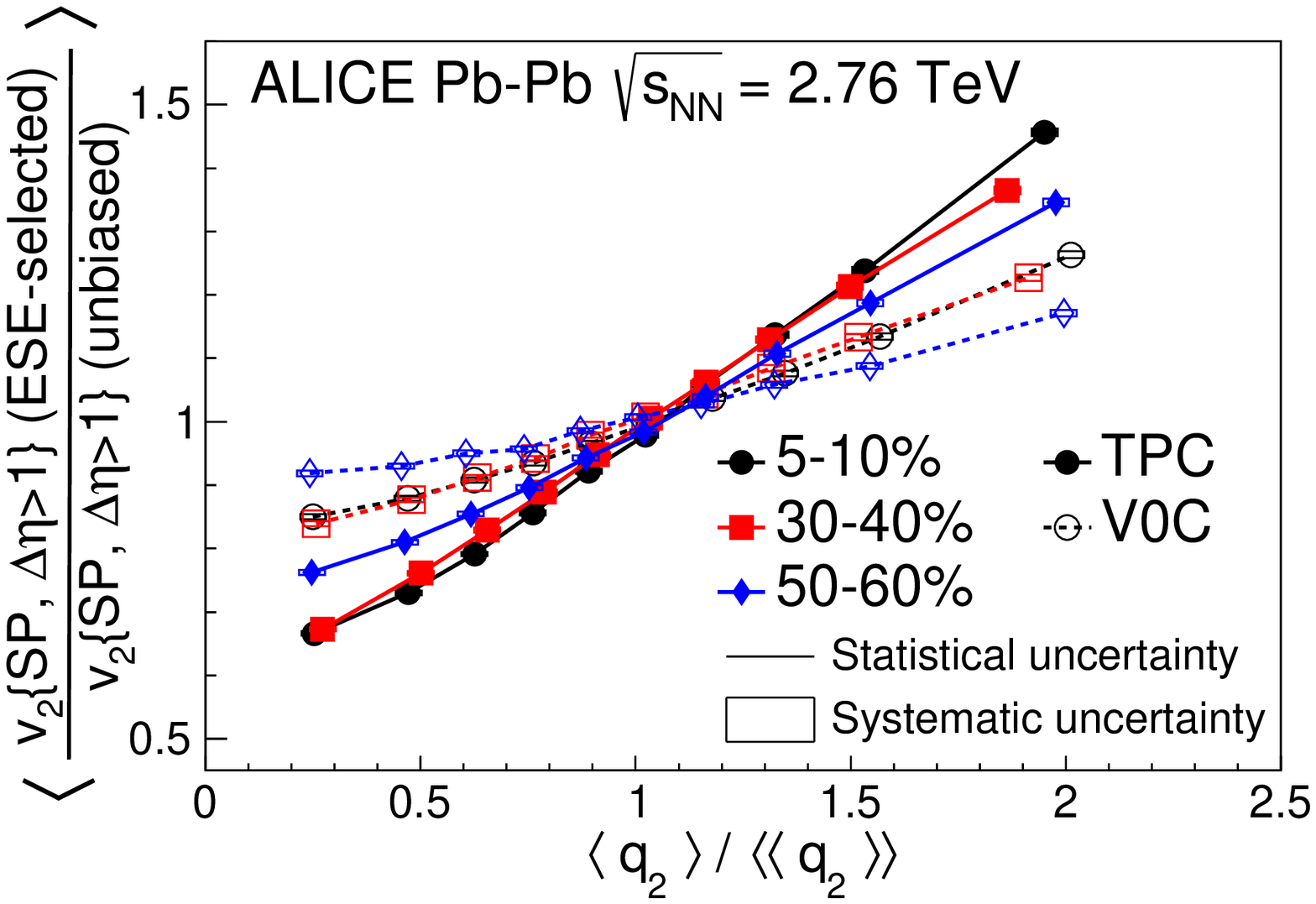}
  \caption{(Color online) Average \vtsp\ variation as a function of the absolute
    (left) values and self-normalized (right) values of the
    \qttpc\ and \qtvc\ for several centrality classes.}
  \label{fig:v2ese-vs-q-AllCent}
\end{figure}
\begin{figure}[tbh!]
  \centering
  \includegraphics[width=\textwidth]{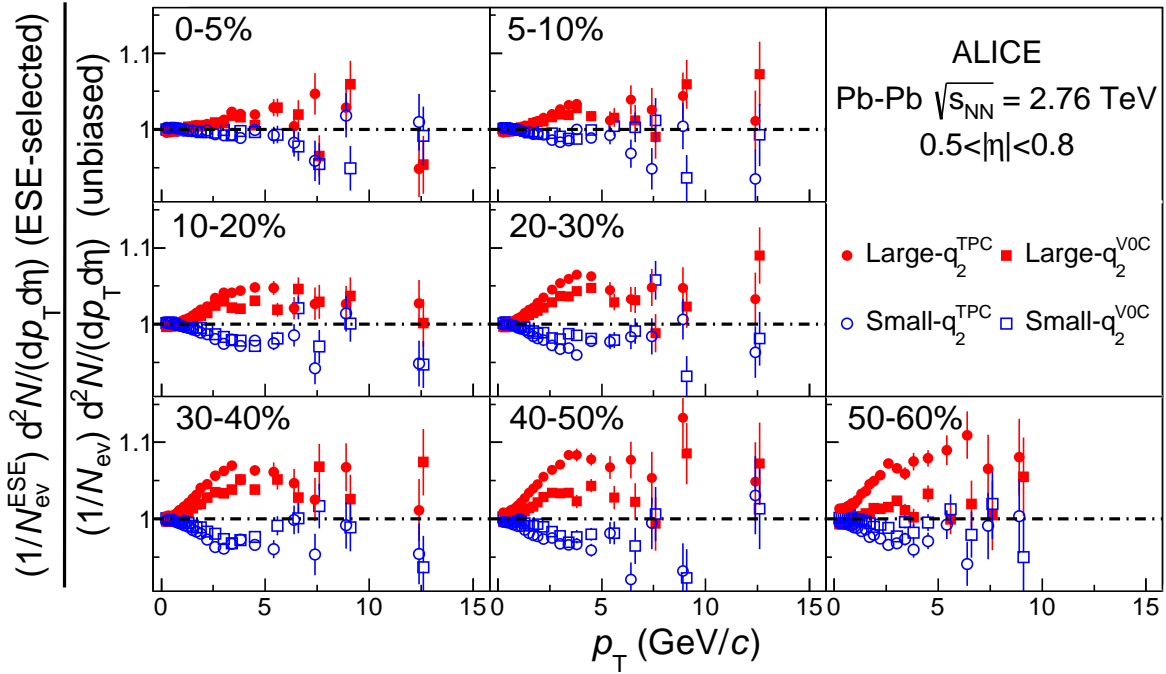}
  \caption{(Color online) Ratio of the \pt\ distribution of charged hadrons in the
    \hq\ or \sq\ sample to the unbiased sample (\qtvc\ and
    \qttpc\ selections) in different centrality classes. Only
    statistical uncertainties are plotted (systematic uncertainties
    are smaller than the markers).}
  \label{fig:spectra-hardons-ese}
\end{figure}

These checks demonstrate that the selectivity of the
cut is the main reason for the difference between the \TPC\ and
\VZEROC\ selections. Due to the large $\eta$ gap, the non-flow
contribution is expected to be negligible in the case of the
\qtvc\ selection. The agreement observed in \Fig{fig:v2ese-relaxed}
indicates that, in the centrality classes 10--50\%, this is also the
case for the \qttpc\ selection in the range $\pt< 5~\gevc$, a
transverse momentum region dominated by hydrodynamic
effects~\cite{Abelev:2013vea}.  It is worth noticing that the ATLAS
Collaboration measured a modification of the elliptic flow of $\sim
35\%$, nearly independent of \pt\ up to $\sim 12~\gevc$ in the
20--30\% centrality class, while measuring $\vtwo$ and $\qtwo$ with a
pseudorapidity gap of 0.7 units~\cite{Aad:2015lwa}. The increasing
trend in the centrality class 0--5\% is also observed
in~\cite{Aad:2015lwa} \footnote{See auxiliary figures available on the
  ATLAS Collaboration web page
  \\\texttt{https://atlas.web.cern.ch/Atlas/GROUPS/PHYSICS/PAPERS/HION-2014-03/}}.

To study the centrality and the \qtwo\ dependence of \vtsp\ in
ESE-selected event classes, we quantified the average change for each
centrality class fitting the ratios in the range $0.2<\pt<4~\gevc$
with a constant\footnote{The result of the fit is numerically
equivalent to the direct computation of the integrated $\vtwo$ in
the range $0.2<\pt<20~\gevc$.}.  The centrality dependence of the
average change in the \hq\ and \sq\ selection is reported in
\Fig{fig:v2ese-vs-centr}.  The trend obtained with the \qttpc\ and
\qtvc\ selections is very similar, except for the most central class
0--5\%, where the average is influenced by the non-flat trend seen in
Fig.~\ref{fig:v2ese-ratio}.  This once again reinforces the conclusion
that the non-flow contamination is small also in the \TPC\ selection
case for the bulk of particles.  The relative importance of non-flow
changes with centrality. A large non-flow bias would therefore
introduce a centrality dependence in the relative trend between the
\qttpc\ and \qtvc\ selections, which is not observed.  The dependence
of the \vtsp\ variation on \qttpc\ and \qtvc\ is shown for the
centrality classes 5-10\%, 30-40\% and 50-60\% in
Fig.~\ref{fig:v2ese-vs-q-AllCent}. The left panel shows the absolute
\qtwo\ values on the x axis, while the right panel depicts the
self-normalized values, defined as the average \qtwo\ value in
ESE-selected events over the average \qtwo\ values for all events in a
given centrality class.  The \VZEROC\ selection spans a larger range
but the \TPC\ is more selective, as is clearly seen from the different
slope of the \TPC\ and \VZEROC\ curves. In both cases the average
\qtwo\ reaches values twice as large compared to those in the unbiased
sample, (Fig.~\ref{fig:v2ese-vs-q-AllCent}, right).

In summary, the observations reported in this section indicate that
the ESE selects a global property of the collisions, as suggested by
the flat modification in the \vtwo\ as a function of \pt. The
\qttpc\ leads to a change twice as large than the corresponding
\qtvc\ selection. The difference between the two seems to be mostly
due to the different discriminating power rather than to non-flow
effects.

\begin{figure}[tbh!]
 \centering
 \includegraphics[width=0.8\textwidth]{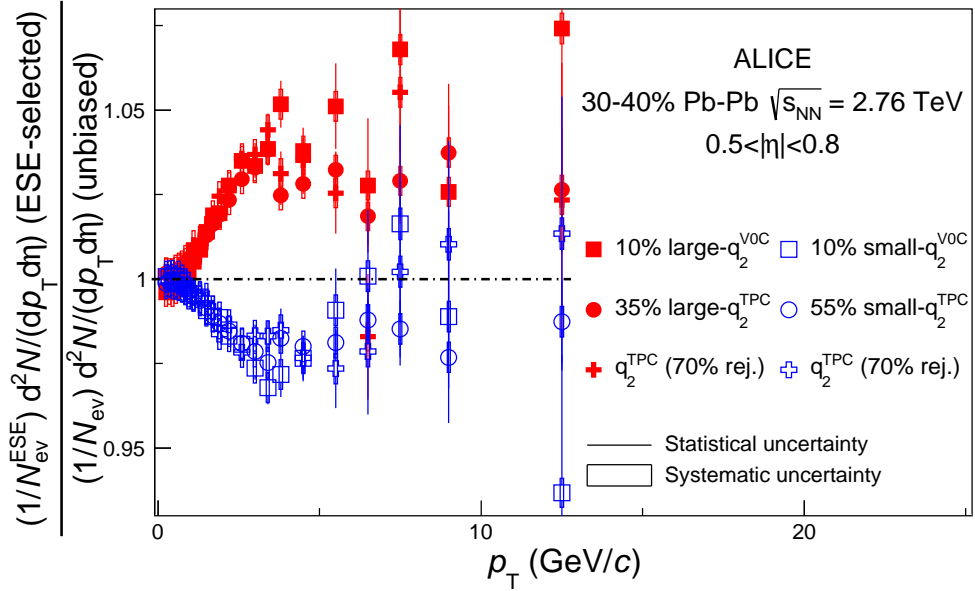}
 \caption{(Color online) Ratio of the \pt\ distribution of charged hadrons in the
 \hq\ or \sq\ sample to the unbiased sample. Comparison between
 the effect of the event-shape selection obtained with the
 standard \VZEROC\ and with the tuned \TPC\ selections (see text
 for details), in the centrality class 30-40\%.}
 \label{fig:spectra-hardons-ese-relaxed}
\end{figure}

\subsection{Transverse momentum distributions}
\label{sec:spectra-results}

In order to study the interplay between the initial configuration of
the system and the dynamics of the expansion of the fireball, the
effect of the ESE selection on the single particle \pt\ distribution
is reported in Fig.~\ref{fig:spectra-hardons-ese}, for the \qttpc\ and
\qtvc\ selections. As discussed in \Sect{sec:event-selection} the
reduced flow vector is calculated in the \TPC\ detector in the
pseudorapidity range $|\eta|<0.4$. In order to avoid overlap between
the \qttpc\ and \pt\ distribution measurements, only the region
$0.5<|\eta|<0.8$ is used to measure the \pt\ distributions.  This
ensures at least 0.1 units of pseudorapidity separation between the
\qtwo\ and spectra measurements, thus suppressing the effect of
short-range correlations. For consistency with the \TPC\ analysis, the
same pseudorapidity range is used in the case of the
\VZEROC\ selection. In the \qtvc\ case, it is also possible to study
the spectra at mid-rapidity $|\eta|<0.8$ without any overlap with the
\qtwo\ measurement. The results agree within uncertainty with those in
$0.5<|\eta|<0.8$.

The spectra in the \hq\ sample are harder than those in the \sq\
one. The ratio to the unbiased spectra reaches a maximum around
$\pt = 4 ~\gevc$, and then stays approximately constant within large
uncertainties.

\begin{figure}[tbh!]
  \centering
  \includegraphics[width=\textwidth]{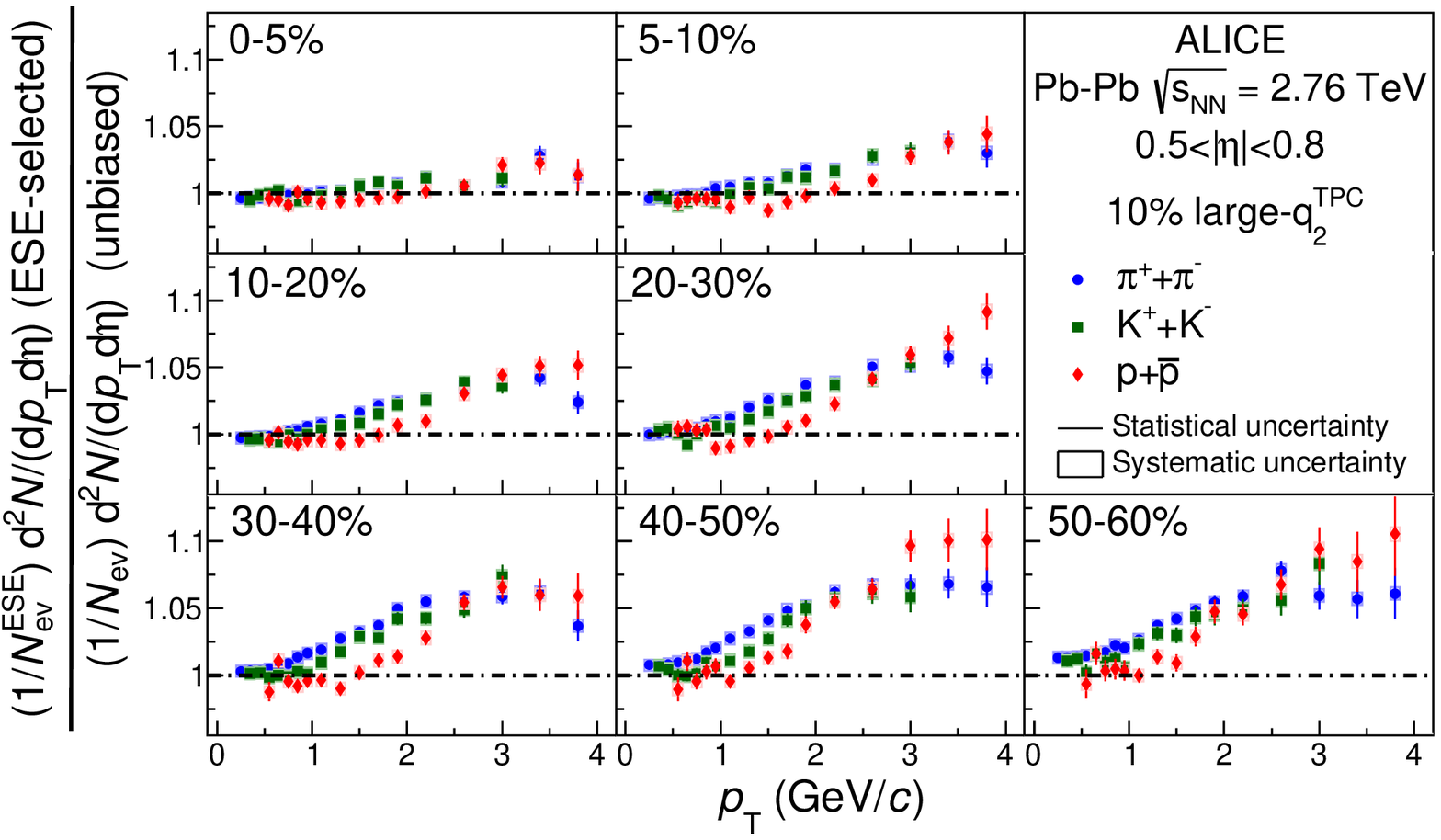}
  \hfill
  \includegraphics[width=\textwidth]{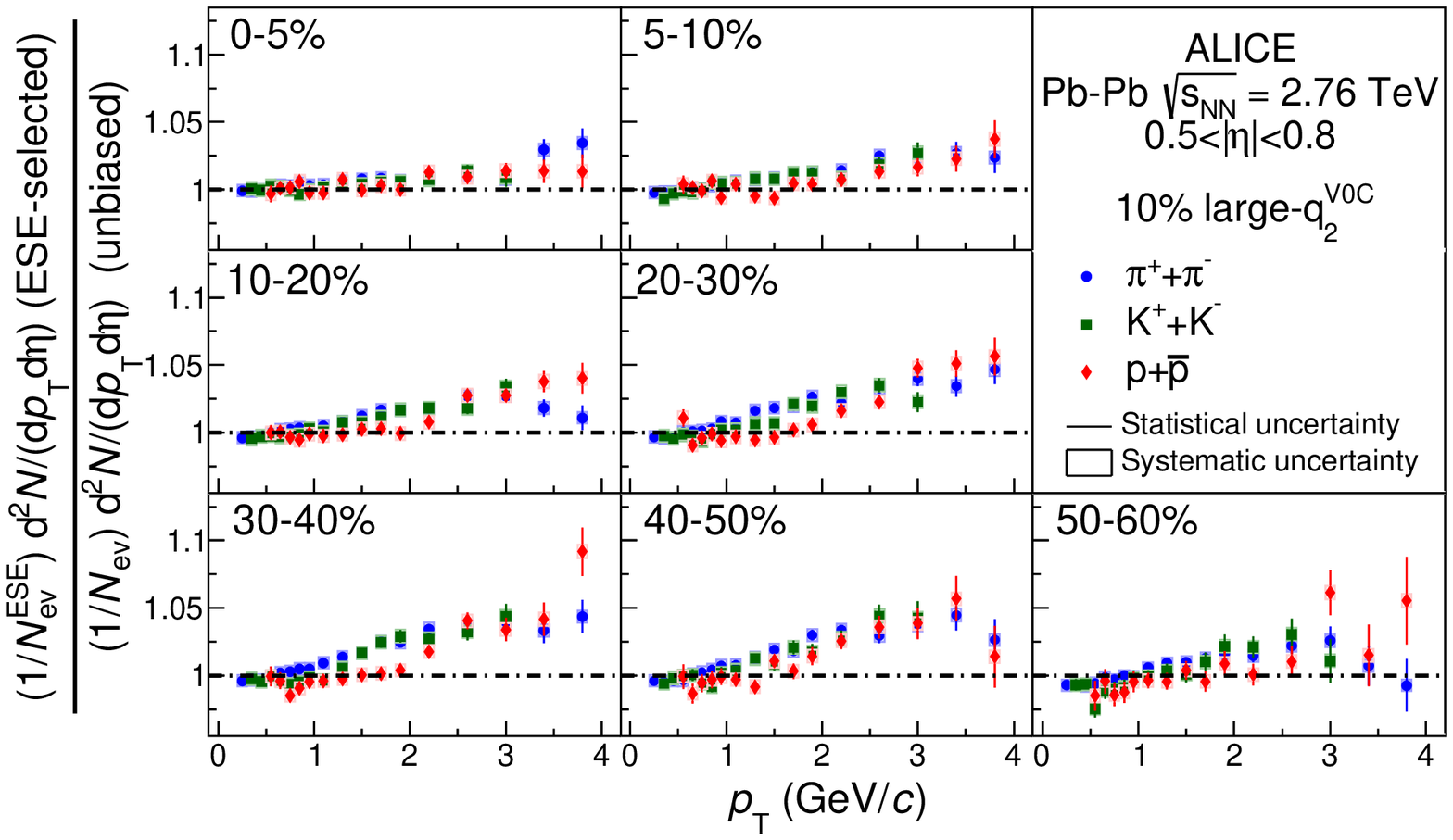}
  \caption{(Color online) Ratio of the \pt\ distribution of identified charged
    hadrons in the \hq\ sample to the unbiased sample for the
    \qttpc\ (top) and \qtvc\ (bottom) selections.}
  \label{fig:spectra-ese-hq}
\end{figure}

\begin{figure}[tbh!]
  \centering
  \includegraphics[width=\textwidth]{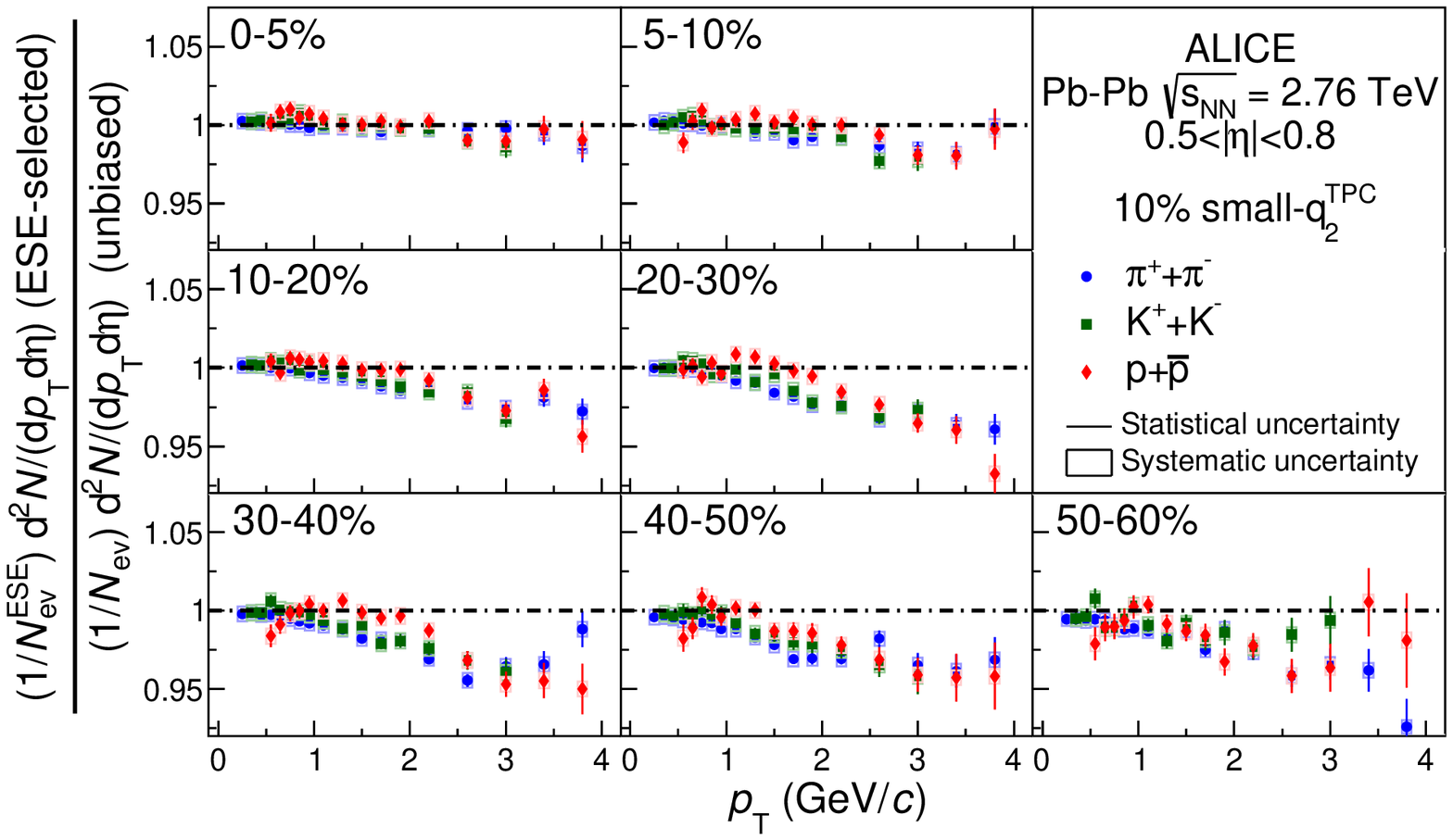}
  \hfill
  \includegraphics[width=\textwidth]{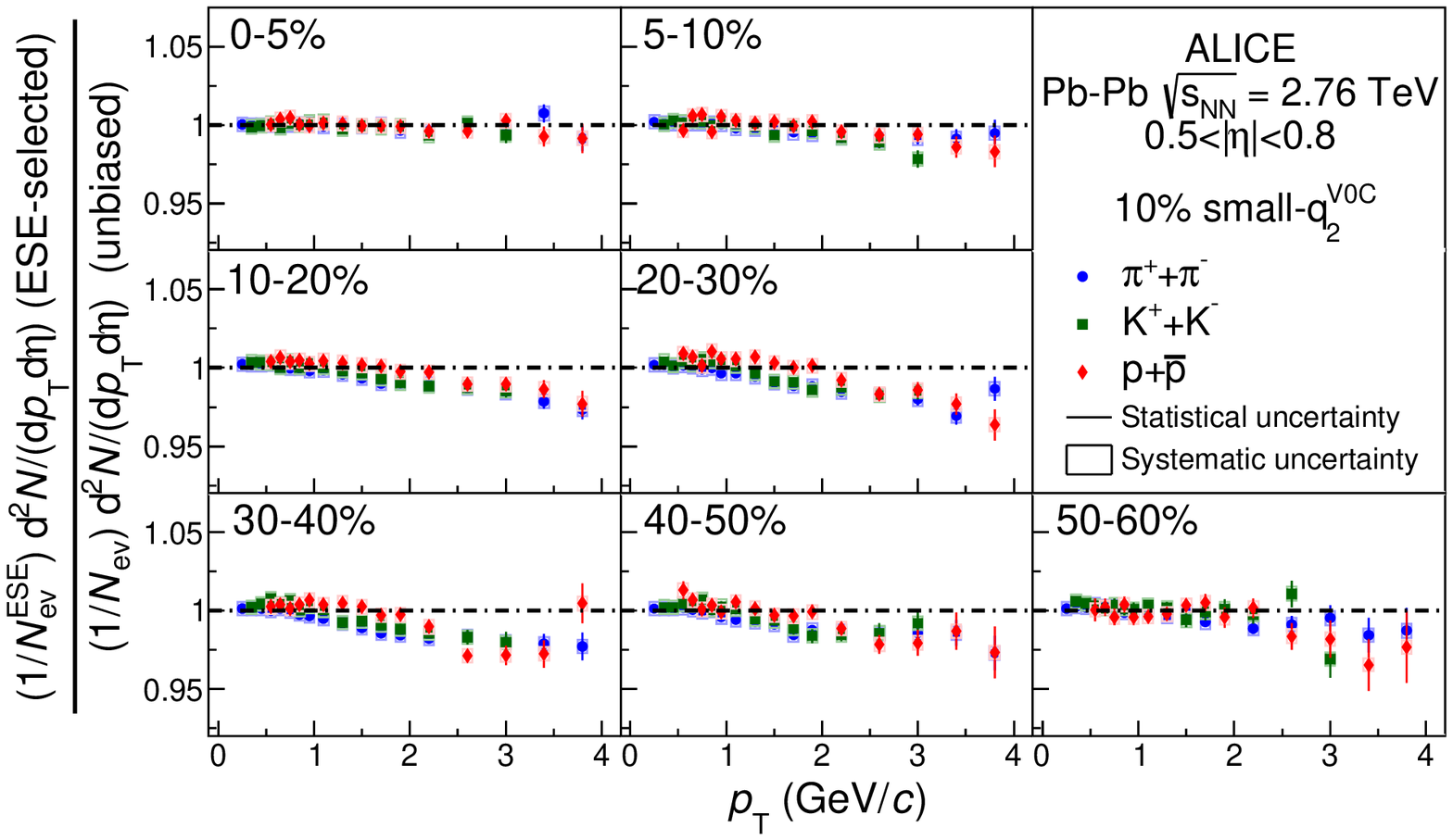}
  \caption{(Color online) Ratio of the \pt\ distribution of identified charged
    hadrons in the \sq\ sample to the unbiased sample for the
    \qttpc\ (top) and \qtvc\ (bottom) selection.}
  \label{fig:spectra-ese-sq}
\end{figure}

The effect of the selection is more pronounced in semi-central events
($\sim$ 30--50\%), and decreases both towards more central and more
peripheral collisions.  This can be understood as due to the fact that
the \qtwo\ spans a larger dynamic range in semi-central collisions
(\Fig{fig:q-vector} and \Fig{fig:v2ese-vs-q-AllCent}).  In the most
peripheral centrality class studied in this paper (50--60\%) the
effect of the TPC-based selection is still very pronounced, while the
\qtvc\ selection is less effective. This may indicate a small
contamination from non-flow effects in the most peripheral bin,
consistent with observations discussed for the \vtsp\ measurement in
\Sect{sec:flow-results}. In the most central class (0-5\%) the
modification of the spectrum is very small.  This suggests that the
trend observed in the same centrality class in \Fig{fig:v2ese-ratio}
is likely to be dominated by flow fluctuations rather than non-flow
contributions.

As in the previous section, we disentangle the effect of non-flow and
\qtwo\ selectivity either relaxing the \qttpc\ selection or randomly
rejecting a fraction of the tracks. The relaxed cut and the fraction
of rejected tracks tuned to reproduce the \vtwo\ variation in
$0.2<\pt<4~\gevc$ in \Sect{sec:flow-results} are used.
\Figure{fig:spectra-hardons-ese-relaxed} shows that these selections
yield results compatible with the standard \qtvc\ selection. A similar
result (with the same relaxed cuts or fraction of rejected tracks) is
found for all centralities up to $\sim$ 50\%, after which non-flow
effects seem to become relevant.

As discussed in
\Sect{sec:flow-results}, we conclude that the effect of non-flow is
small and that the main factor driving these observations is the
average \vtwo\ at mid-rapidity.

\begin{figure}[t!]
  \centering
  \includegraphics[width=0.7\textwidth]{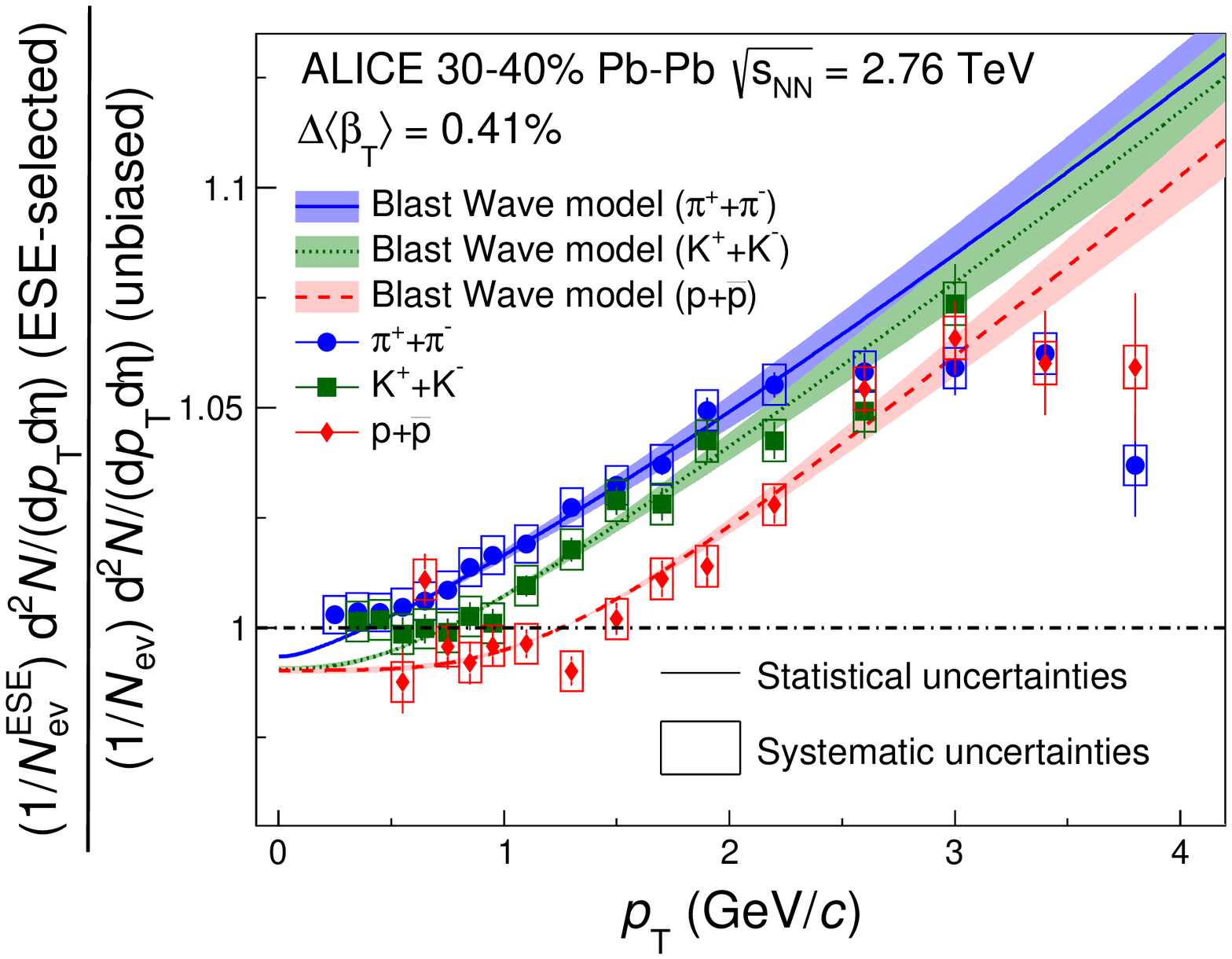}
  \hfill
  \includegraphics[width=0.7\textwidth]{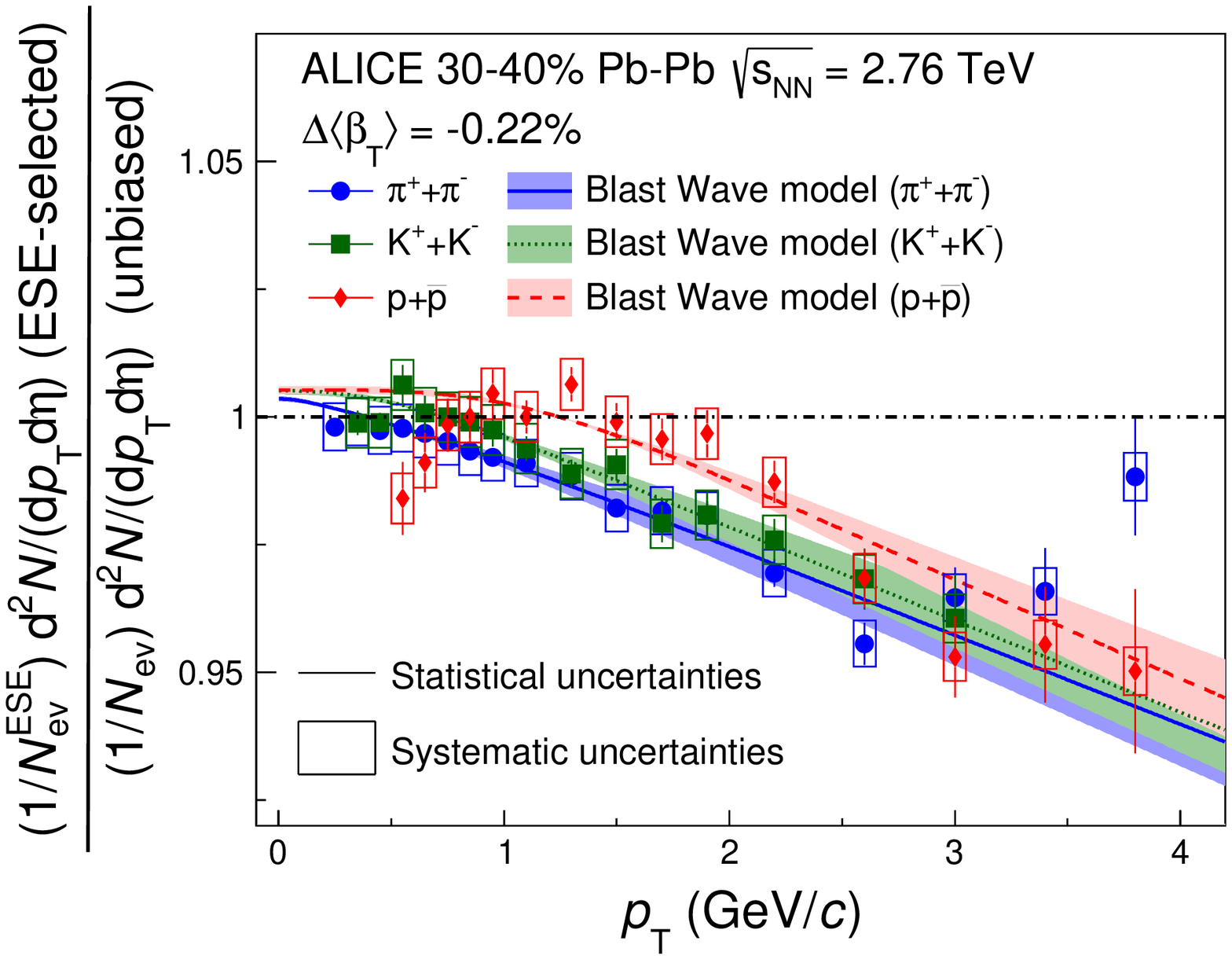}
  \caption{(Color online) Ratio of the \pt\ distribution of identified charged hadrons in the \hq\ (top) and \sq\ (bottom) sample to the unbiased sample (\qttpc\ selection), in 30-40\% centrality class. Lines: ratio of the blast-wave parametrizations (see text for details).}
  \label{fig:spectra-ese_BW}
\end{figure}

The modification on the spectra of identified $\pi$, K, and p is
reported in \Fig{fig:spectra-ese-hq} and \Fig{fig:spectra-ese-sq} for
different centrality classes. The same pattern measured in the case of
non-identified hadrons is observed. Moreover, a clear mass ordering is
seen: the modification is more pronounced for heavier particles.
Conversely, the spectra in the \sq\ sample are softer.  In the case of
the \VZEROC\ selection the analysis was also repeated in the region
$|y|<0.5$, yielding consistent results.
 
These observations suggest that the spectra in the \hq\ (\sq) sample
are affected by a larger (smaller) radial flow push. This hypothesis
was tested with a blast-wave~\cite{Schnedermann:1993ws} study.  A
ratio of two blast-wave functions was used to fit the spectra ratios
shown in \Fig{fig:spectra-ese-hq} and \Fig{fig:spectra-ese-sq}. The
parameters were initially fixed to the values
from~\cite{Abelev:2013vea}, where they were tuned to describe the
inclusive spectra of pions, kaons and protons. Then, the
\avbT\ parameter of the numerator function was allowed to change
(while keeping the overall integral of the function constant). The fit
was performed as in~\cite{Abelev:2013vea} in the transverse momentum
ranges 0.5-1 \gevc, 0.2-1.5 \gevc, 0.3-3 \gevc\ for $\pi$, K, p,
respectively. The agreement with the data is good, also outside the
range used to determine the parameters, up to $\pt\sim 3 \gevc$. The
fits yield the following result for the difference $\Delta\avbT$
between the \avbT\ parameter of the numerator and denominator
function: $\Delta\avbT = (0.41\pm0.03)$\% (\hq) and $\Delta\avbT =
(-0.22\pm0.03)$\% (\sq) for the centrality class 30-40\%, as shown in
Fig.~\ref{fig:spectra-ese_BW}.


\section{Discussion}
\label{sec:discussion}
In this paper the first application of the Event Shape Engineering
(ESE) \cite{Schukraft:2012ah} to the analysis of ALICE data was
presented.

The results on the \vtsp\ measurement suggest that the ESE technique
selects a global property of the collision, likely related to the
eccentricity in the initial state. The measurement of \pt\ spectra
indicates that events with larger eccentricity show an increased
radial flow.  A correlation between elliptic and radial flow could be
introduced either at the initial stage, due to the specific
fluctuation patterns in the energy deposition, or during the
hydrodynamic evolution of the system, due to an interplay of bulk and
shear viscosity~\cite{Heinz:2013th}.

\begin{figure}[t!]
  \centering
  \includegraphics[width=0.6\textwidth]{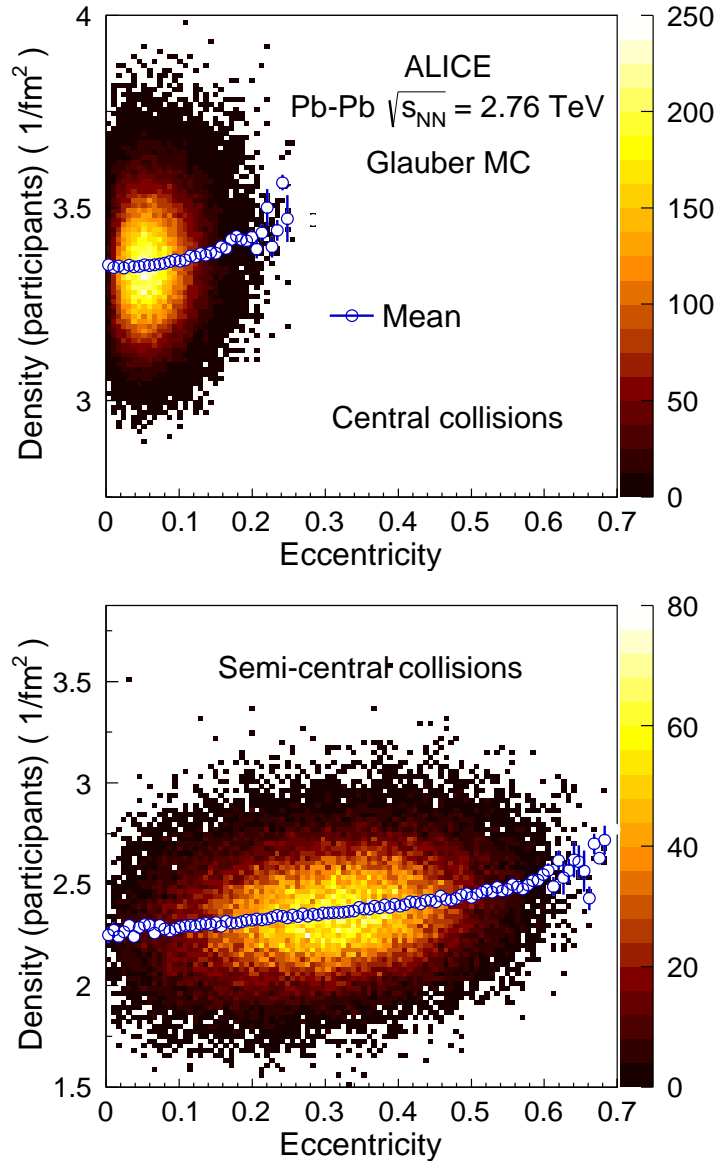}
  \caption{(Color online) Participant density as a function of the participant
    eccentricity estimated in a Glauber Monte Carlo model for central
    (top) and semi-central (bottom) collisions.}
  \label{fig:glauber}
\end{figure}

A Glauber Monte Carlo simulation was performed to estimate the
possible correlation between the initial eccentricity and
azimuthally-averaged pressure gradients. In the model, the
multiplicity of charged particles in the acceptance of the
\VZERO\ detector, used to determine the centrality classes, is
computed following~\cite{Abelev:2013qoq}. A ``number of ancestors''
$N_{ancestors}$ is derived from the number of participant nucleons
($N_{part}$) and binary collisions ($N_{coll}$) as
\begin{equation}
  \label{eq:ancestor}
  N_{ancestors} = f  N_{part}+(1 - f ) N_{coll}.
\end{equation}
Each ancestor is assumed to produce particles following a negative
binomial distribution with parameters taken
from~\cite{Abelev:2013qoq}.

The participant density, defined
following~\cite{Voloshin:1999gs,Alver:2008zza,Alver:2006wh,Voloshin:2007pc}
as $N_{part}/S$, is used as a proxy for the magnitude of the pressure
gradients. The average cross-sectional area $S$ and participant
eccentricity $\epsilon$ are computed as

\begin{equation}
  \label{eq:area} S =4 \pi \sigma_{x'}
  \sigma_{y'} =4\pi \sqrt{\sigma_x^2 \sigma_y^2 -\sigma_{xy}^2},
\end{equation}
\begin{equation}
  \label{eq:eccentricity}
  \epsilon=\frac{\sigma_{y'}^{2}-\sigma_{x'}^{2}}
  {\sigma_{x'}^{2}+\sigma_{y'}^{2}}
  =\frac{\sqrt{(\sigma_y^2-\sigma_x^2)^2+4\sigma_{xy}^2}}{\sigma_x^2+\sigma_y^2},
\end{equation} where
\begin{equation}
  \label{eq:sigma} \sigma_{x}^{2} =\langle x^{2}\rangle
  - \langle x \rangle^{2},\;\; \sigma_{y}^{2} =\langle y^{2}\rangle -
  \langle y \rangle^{2},\;\; \sigma_{xy} =\langle xy\rangle - \langle x
  \rangle \langle y \rangle.
\end{equation} 

The unprimed coordinates are given in the fixed
laboratory coordinate frame.  
Primed coordinates, $x'$ and $y'$, are calculated in the so-called
participant coordinate system, rotated with respect to the 
laboratory coordinate frame 
such that the minor symmetry axis of
the participant nucleon distribution coincides with the $x'$
direction. 
The normalization of the area is chosen such that
for a Gaussian distribution the {\em average} density coincides with
$N_{part}/S$.

Two narrow centrality classes, selected based on the simulated charged
particle multiplicity, roughly corresponding to 0-2\% (central) and
30-32\% (semi-central), are studied in Fig.~\ref{fig:glauber}. The
observed correlation between the density and the participant
eccentricity is reminiscent of the correlation between radial flow and
event shape measured in this paper. The average density in events with
the 10\% largest $\epsilon$ is about 1\% (7\%) larger than in events
with the smallest $\epsilon$ for central (semi-central) collisions,
qualitatively consistent with what is observed in
\Fig{fig:spectra-ese-hq} and \Fig{fig:spectra-ese-sq}, where the
effect of the ESE selection is much stronger for semi-central
collisions.  This reinforces our conclusion that ESE is an effective
tool to select the initial shape and density, thereby opening the
possibility of further studies.

A quantitative comparison would require a full hydrodynamical
calculation.  The correlation can in fact be modified by the transport
in the hydrodynamic phase. In particular, it was
shown~\cite{Heinz:2013th,Song:2009rh} that in a system with a finite
shear viscosity the flow coefficients, obtained for a given set of
initial eccentricities, are reduced as compared to the ideal
hydrodynamics case. At the same time, shear viscosity increases the
radial flow. In principle, bulk viscosity reduces the radial flow,
reducing the correlation observed in this paper, but the latter effect
was estimated to be negligible~\cite{Song:2009rh}.  Therefore, the
measurement we present in this paper is sensitive to the interplay of
initial conditions and transport coefficients in the hydrodynamic
phase. As such, it poses stringent constraints on hydrodynamic
calculations, and it could allow the extraction of the value of
average shear viscosity at the LHC.

\begin{figure}[t!]
  \centering
  \includegraphics[width=0.75\textwidth]{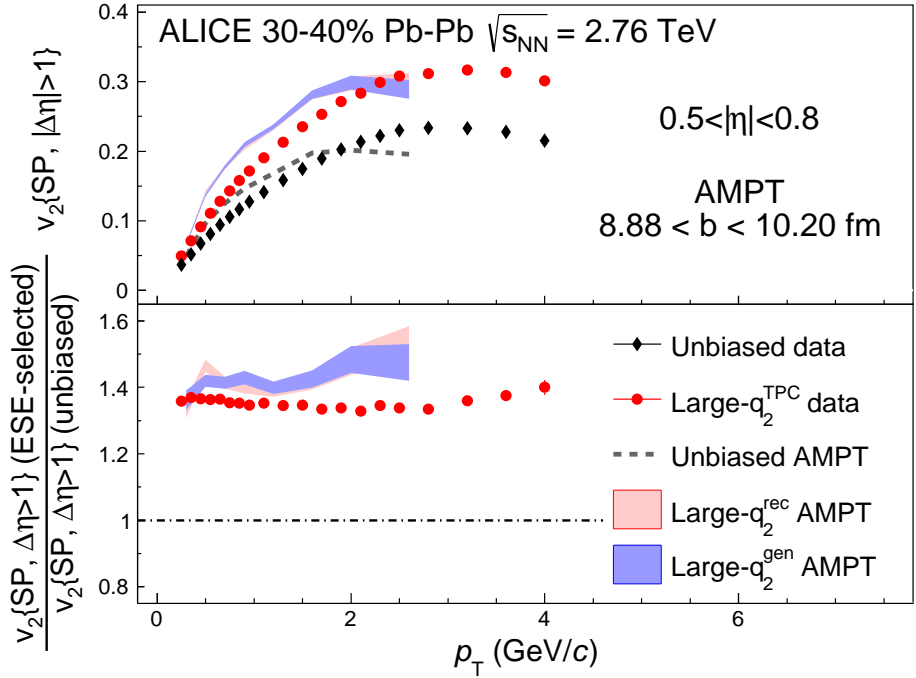}
   \caption{(Color online) Measurement of \vtsp\ as a function of \pt, for the unbiased sample and
     for the \hq\ sample (top panel) and ratio between the \hq\ result
     over the unbiased result (bottom panel). Data points (full
     markers) are compared with AMPT Monte Carlo model (bands). Only
     statistical uncertainties are plotted (systematic uncertainties
     are smaller than the markers).}
  \label{fig:ampt-ese-v2}
\end{figure}

\begin{figure}[t!]
  \centering
  \includegraphics[width=0.75\textwidth]{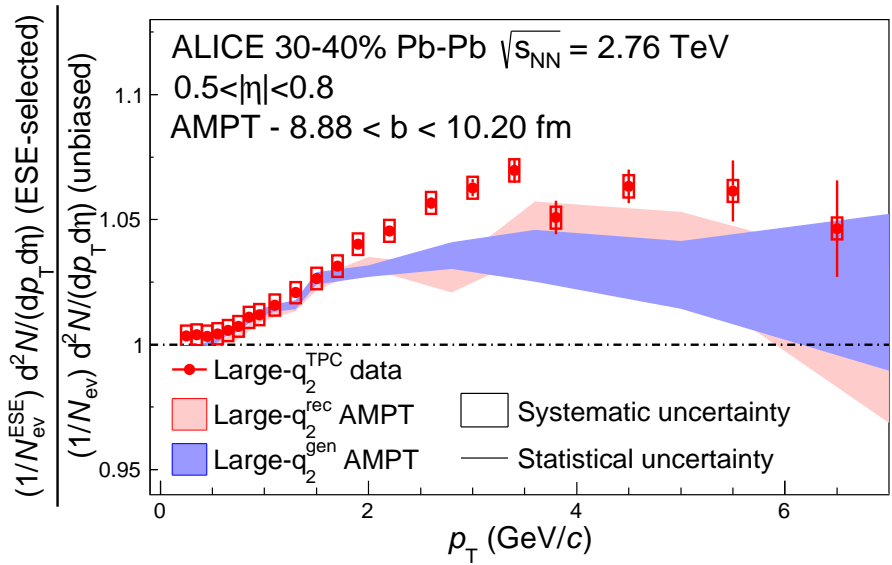}
  \caption{(Color online) Ratio of the \pt\ distribution of charged hadrons in the
    \hq\ sample to the unbiased sample for the \qttpc\ selection. Data
    points (full markers) are compared with AMPT Monte Carlo model
    (bands).}
  \label{fig:ampt-ese-spectra}
\end{figure}

A study of the relation of the fluctuation in the initial size to the
spectra was performed in~\cite{Bozek:2012fw,Broniowski:2009fm} with a
full hydrodynamic simulation. It was shown that the event-by-event
fluctuations in the Glauber initial conditions lead to fluctuations in
the initial size of the system that reflect in fluctuations of the
radial flow and hence \avpT. It is found that the relative \avpT\
fluctuations computed with Glauber initial conditions overestimate the
data, indicating a strong sensitivity of event-by-event measurements
on the initial conditions model. It is also shown that the \avpT\
fluctuations are not sensitive to the shear viscosity. The study
in~\cite{Bozek:2012fw,Broniowski:2009fm} (fluctuations in \avpT),
however, does not address the relation between the elliptic and radial
flow. It may be expected that the present measurement will also be
sensitive to the transport coefficient of the medium.

In a recent series of theoretical
studies~\cite{Bhalerao:2014mua,Mazeliauskas:2015vea,Mazeliauskas:2015efa},
it was suggested to use the Principal Component Analysis to study flow
fluctuations. It was argued that most of the current methods to study
flow do not fully capture the complexity of the initial state. Indeed,
the PCA studies revealed the presence of “sub-leading” flow components
(arising from radial geometry excitations), which break the
factorization of flow
harmonics~\cite{Bhalerao:2014mua,Mazeliauskas:2015vea}, In particular,
in~\cite{Mazeliauskas:2015efa} it is argued that the sub-leading
component of \vtwo\ reflects a non linear mixing with radial flow,
which could address the same physics as reported in this paper.

A study of the relation of the fluctuation in the initial size to the
spectra was performed in~\cite{Bozek:2012fw,Broniowski:2009fm} with a
full hydrodynamic simulation. It was shown that the event-by-event
fluctuations in the Glauber initial conditions lead to fluctuations in
the initial size of the system that reflect in fluctuations of the
radial flow and hence \avpT. It is found that the relative \avpT\
fluctuations computed with Glauber initial conditions overestimate the
data, indicating a strong sensitivity of event-by-event measurements
on the initial conditions model. It is also shown that the \avpT\
fluctuations are not sensitive to the shear viscosity. The study
in~\cite{Bozek:2012fw,Broniowski:2009fm} (fluctuations in \avpT),
however, does not address the relation between the elliptic and radial
flow. It may be expected that the present measurement will also be
sensitive to the transport coefficient of the medium.

In a recent series of theoretical
studies~\cite{Bhalerao:2014mua,Mazeliauskas:2015vea,Mazeliauskas:2015efa},
it was suggested to use the Principal Component Analysis to study flow
fluctuations. It was argued that most of the current methods to study
flow do not fully capture the complexity of the initial state. Indeed,
the PCA studies revealed the presence of “sub-leading” flow components
(arising from radial geometry excitations), which break the
factorization of flow
harmonics~\cite{Bhalerao:2014mua,Mazeliauskas:2015vea}. In particular,
in~\cite{Mazeliauskas:2015efa} it is argued that the sub-leading
component of \vtwo\ reflects a non linear mixing with radial flow,
which could be related to the same underlying physics phenomena
reported in this paper.

To further understand the observed effect, we studied it in AMPT, a model
known to reproduce many of the flow observables measured at the
LHC~\cite{Lin:2004en}. This model is based on HIJING
\cite{Gyulassy:1994ew} to describe the initial conditions and on the
Zhang's parton cascade \cite{Zhang:1997ej} to describe the partonic
evolution. The \textit{string melting} configuration, described in
\cite{Abelev:2013csa}, is used. To assess the impact of the detector
resolution on the \qtwo\ selection, the simulated AMPT events were
transported through the ALICE apparatus using the
GEANT~\cite{Geant:1994zzo} transport model.  The \qtwo\ was computed
using either the reconstructed Monte Carlo tracks (\qtrec) or the
generated primary particles in the same kinematic range (\qtgen). The
elliptic flow and the transverse momentum distribution are calculated
using generated Monte Carlo particles.  Since the charged particle
multiplicity distribution is different in AMPT and data, the \qtwo\
selection is calibrated in the model as a function of multiplicity.
The results are shown in \Fig{fig:ampt-ese-v2} for the charged hadrons
elliptic flow and in \Fig{fig:ampt-ese-spectra} for the transverse
momentum distribution of charged hadrons. Using either \qtrec\ or
\qtgen\ does not introduce any significant difference on the effect of
the selection. This indicates that detector resolution effects are
negligible for the \qttpc\ selection.  The \VZERO\ detectors, on the
other hand, have a coarser azimuthal resolution and are sensitive to
fluctuations in the energy deposition of incident particles.  However,
the study with the relaxed TPC selection discussed in
\Sect{sec:results} demonstrates that the properties of the ESE
selected events are mostly determined by the average \vtsp\ value.  It
is therefore advised that in any comparison of this data to
theoretical models the selection in the model is tuned as to reproduce
the average change in \vtsp\ at mid-rapidity.

The \pt\ dependence of the elliptic flow observed in data is not
reproduced in AMPT (top panel). This model reproduces however the
magnitude of the modification as well as the flatness of the ratio as
a function of \pt.

The effect of the ESE selection on the \pt\ distribution of charged
particles is well reproduced by AMPT below $\pt~=~2~\gevc$, as shown
in \Fig{fig:ampt-ese-spectra}. However, the magnitude of the effect at
intermediate \pt\ ($2<\pt<6~\gevc$) is underestimated in AMPT. As
previously observed for the \vtwo\ measurement, a good agreement is
observed between the selection based on \qtgen\ and \qtrec.

\section{Conclusions}
\label{sec:conclusion}

In summary, the first application of the Event Shape Engineering (ESE)
technique to \PbPb\ collisions data measured by ALICE at
\snn~=~2.76~TeV has been presented.

The elliptic flow at mid-rapidity is observed to increase as a
function of the \qtwo\ calculated in the central or forward rapidity
regions. The modification of the \vtwo\ coefficient as a function of
\pt\ is nearly flat below $\pt~=~4$~\gevc, suggesting that this
technique allows the selection of a global property of the collision,
likely related with the geometry of the participant nucleons in the
initial state. In the region above $\pt>5$~\gevc\ a small increase is
observed within the large statistical uncertainties, possibly due to a
small non-flow contamination.  In this transverse momentum range the
elliptic flow is believed to be driven by the different path-length
traversed in- and out-of-plane by high-\pt\ partons in the deconfined
medium, rather than by the hydrodynamic evolution of the system.

The \pt-distributions of unidentified hadrons in the \pt\ region
($0<\pt<5$~\gevc) are harder (softer) in event with \hq\ (\sq) values.

\enlargethispage{0.5cm}
Identified pions, kaons and protons show a similar behavior with a
clear mass ordering in the ratio between the \hq\ and the unbiased
spectra, thus suggesting this effect to be due to a stronger radial
flow in such events.
Glauber Monte Carlo calculations reveal a correlation between the
transverse participant density and the participant eccentricity which
could be the origin of this effect. This indicates that at least part
of the correlation is generated in the initial state. However, these
measurements are also sensitive to the transport coefficients in the
hydrodynamic evolution. A quantitative comparison would require a full
hydrodynamic calculation and may provide stringent constraints both on
shear and bulk viscosity.


%% file: acknowledgements.tex

The ALICE Collaboration would like to thank all its engineers and technicians for their invaluable contributions to the construction of the experiment and the CERN accelerator teams for the outstanding performance of the LHC complex.
The ALICE Collaboration gratefully acknowledges the resources and support provided by all Grid centres and the Worldwide LHC Computing Grid (WLCG) collaboration.
The ALICE Collaboration acknowledges the following funding agencies for their support in building and
running the ALICE detector:
State Committee of Science,  World Federation of Scientists (WFS)
and Swiss Fonds Kidagan, Armenia,
Conselho Nacional de Desenvolvimento Cient\'{\i}fico e Tecnol\'{o}gico (CNPq), Financiadora de Estudos e Projetos (FINEP),
Funda\c{c}\~{a}o de Amparo \`{a} Pesquisa do Estado de S\~{a}o Paulo (FAPESP);
National Natural Science Foundation of China (NSFC), the Chinese Ministry of Education (CMOE)
and the Ministry of Science and Technology of China (MSTC);
Ministry of Education and Youth of the Czech Republic;
Danish Natural Science Research Council, the Carlsberg Foundation and the Danish National Research Foundation;
The European Research Council under the European Community's Seventh Framework Programme;
Helsinki Institute of Physics and the Academy of Finland;
French CNRS-IN2P3, the `Region Pays de Loire', `Region Alsace', `Region Auvergne' and CEA, France;
German Bundesministerium fur Bildung, Wissenschaft, Forschung und Technologie (BMBF) and the Helmholtz Association;
General Secretariat for Research and Technology, Ministry of
Development, Greece;
Hungarian Orszagos Tudomanyos Kutatasi Alappgrammok (OTKA) and National Office for Research and Technology (NKTH);
Department of Atomic Energy and Department of Science and Technology of the Government of India;
Istituto Nazionale di Fisica Nucleare (INFN) and Centro Fermi -
Museo Storico della Fisica e Centro Studi e Ricerche "Enrico
Fermi", Italy;
MEXT Grant-in-Aid for Specially Promoted Research, Ja\-pan;
Joint Institute for Nuclear Research, Dubna;
National Research Foundation of Korea (NRF);
Consejo Nacional de Cienca y Tecnologia (CONACYT), Direccion General de Asuntos del Personal Academico(DGAPA), M\'{e}xico, Amerique Latine Formation academique - European Commission~(ALFA-EC) and the EPLANET Program~(European Particle Physics Latin American Network);
Stichting voor Fundamenteel Onderzoek der Materie (FOM) and the Nederlandse Organisatie voor Wetenschappelijk Onderzoek (NWO), Netherlands;
Research Council of Norway (NFR);
National Science Centre, Poland;
Ministry of National Education/Institute for Atomic Physics and National Council of Scientific Research in Higher Education~(CNCSI-UEFISCDI), Romania;
Ministry of Education and Science of Russian Federation, Russian
Academy of Sciences, Russian Federal Agency of Atomic Energy,
Russian Federal Agency for Science and Innovations and The Russian
Foundation for Basic Research;
Ministry of Education of Slovakia;
Department of Science and Technology, South Africa;
Centro de Investigaciones Energeticas, Medioambientales y Tecnologicas (CIEMAT), E-Infrastructure shared between Europe and Latin America (EELA), Ministerio de Econom\'{i}a y Competitividad (MINECO) of Spain, Xunta de Galicia (Conseller\'{\i}a de Educaci\'{o}n),
Centro de Aplicaciones Tecnológicas y Desarrollo Nuclear (CEA\-DEN), Cubaenerg\'{\i}a, Cuba, and IAEA (International Atomic Energy Agency);
Swedish Research Council (VR) and Knut $\&$ Alice Wallenberg
Foundation (KAW);
Ukraine Ministry of Education and Science;
United Kingdom Science and Technology Facilities Council (STFC);
The United States Department of Energy, the United States National
Science Foundation, the State of Texas, and the State of Ohio;
Ministry of Science, Education and Sports of Croatia and  Unity through Knowledge Fund, Croatia.
Council of Scientific and Industrial Research (CSIR), New Delhi, India

%% file: Alice_Authorlist_2015-Jun-23.tex


\begingroup
\small
\begin{flushleft}
J.~Adam\Irefn{org40}\And
D.~Adamov\'{a}\Irefn{org83}\And
M.M.~Aggarwal\Irefn{org87}\And
G.~Aglieri Rinella\Irefn{org36}\And
M.~Agnello\Irefn{org111}\And
N.~Agrawal\Irefn{org48}\And
Z.~Ahammed\Irefn{org132}\And
S.U.~Ahn\Irefn{org68}\And
I.~Aimo\Irefn{org94}\textsuperscript{,}\Irefn{org111}\And
S.~Aiola\Irefn{org136}
% ...
\renewcommand\labelenumi{\textsuperscript{\theenumi}~}

\section*{Affiliation notes}
\renewcommand\theenumi{\roman{enumi}}
\begin{Authlist}
\item \Adef{0}Deceased
\item \Adef{idp3813504}{Also at: M.V. Lomonosov Moscow State University, D.V. Skobeltsyn Institute of Nuclear, Physics, Moscow, Russia}
\item \Adef{idp5960816}{Also at: University of Kansas, Lawrence, Kansas, United States}
\end{Authlist}

\section*{Collaboration Institutes}
\renewcommand\theenumi{\arabic{enumi}~}
\begin{Authlist}

\item \Idef{org1}A.I. Alikhanyan National Science Laboratory (Yerevan Physics Institute) Foundation, Yerevan, Armenia
\item \Idef{org2}Benem\'{e}rita Universidad Aut\'{o}noma de Puebla, Puebla, Mexico
\item \Idef{org3}Bogolyubov Institute for Theoretical Physics, Kiev, Ukraine
\item \Idef{org4}Bose Institute, Department of Physics and Centre for Astroparticle Physics and Space Science (CAPSS), Kolkata, India
\item \Idef{org5}Budker Institute for Nuclear Physics, Novosibirsk, Russia
\item \Idef{org6}California Polytechnic State University, San Luis Obispo, California, United States
\item \Idef{org7}Central China Normal University, Wuhan, China
\item \Idef{org8}Centre de Calcul de l'IN2P3, Villeurbanne, France
\item \Idef{org9}Centro de Aplicaciones Tecnol\'{o}gicas y Desarrollo Nuclear (CEADEN), Havana, Cuba
\item \Idef{org10}Centro de Investigaciones Energ\'{e}ticas Medioambientales y Tecnol\'{o}gicas (CIEMAT), Madrid, Spain
\item \Idef{org11}Centro de Investigaci\'{o}n y de Estudios Avanzados (CINVESTAV), Mexico City and M\'{e}rida, Mexico
\item \Idef{org12}Centro Fermi - Museo Storico della Fisica e Centro Studi e Ricerche ``Enrico Fermi'', Rome, Italy
\item \Idef{org13}Chicago State University, Chicago, Illinois, USA
\item \Idef{org14}China Institute of Atomic Energy, Beijing, China
\item \Idef{org15}Commissariat \`{a} l'Energie Atomique, IRFU, Saclay, France
\item \Idef{org16}COMSATS Institute of Information Technology (CIIT), Islamabad, Pakistan
\item \Idef{org17}Departamento de F\'{\i}sica de Part\'{\i}culas and IGFAE, Universidad de Santiago de Compostela, Santiago de Compostela, Spain
\item \Idef{org18}Department of Physics and Technology, University of Bergen, Bergen, Norway
\item \Idef{org19}Department of Physics, Aligarh Muslim University, Aligarh, India
\item \Idef{org20}Department of Physics, Ohio State University, Columbus, Ohio, United States
\item \Idef{org21}Department of Physics, Sejong University, Seoul, South Korea
\item \Idef{org22}Department of Physics, University of Oslo, Oslo, Norway
\item \Idef{org23}Dipartimento di Elettrotecnica ed Elettronica del Politecnico, Bari, Italy
\item \Idef{org24}Dipartimento di Fisica dell'Universit\`{a} 'La Sapienza' and Sezione INFN Rome, Italy
\item \Idef{org25}Dipartimento di Fisica dell'Universit\`{a} and Sezione INFN, Cagliari, Italy
\item \Idef{org26}Dipartimento di Fisica dell'Universit\`{a} and Sezione INFN, Trieste, Italy
\item \Idef{org27}Dipartimento di Fisica dell'Universit\`{a} and Sezione INFN, Turin, Italy
\item \Idef{org28}Dipartimento di Fisica e Astronomia dell'Universit\`{a} and Sezione INFN, Bologna, Italy
\item \Idef{org29}Dipartimento di Fisica e Astronomia dell'Universit\`{a} and Sezione INFN, Catania, Italy
\item \Idef{org30}Dipartimento di Fisica e Astronomia dell'Universit\`{a} and Sezione INFN, Padova, Italy
\item \Idef{org31}Dipartimento di Fisica `E.R.~Caianiello' dell'Universit\`{a} and Gruppo Collegato INFN, Salerno, Italy
\item \Idef{org32}Dipartimento di Scienze e Innovazione Tecnologica dell'Universit\`{a} del  Piemonte Orientale and Gruppo Collegato INFN, Alessandria, Italy
\item \Idef{org33}Dipartimento Interateneo di Fisica `M.~Merlin' and Sezione INFN, Bari, Italy
\item \Idef{org34}Division of Experimental High Energy Physics, University of Lund, Lund, Sweden
\item \Idef{org35}Eberhard Karls Universit\"{a}t T\"{u}bingen, T\"{u}bingen, Germany
\item \Idef{org36}European Organization for Nuclear Research (CERN), Geneva, Switzerland
\item \Idef{org37}Excellence Cluster Universe, Technische Universit\"{a}t M\"{u}nchen, Munich, Germany
\item \Idef{org38}Faculty of Engineering, Bergen University College, Bergen, Norway
\item \Idef{org39}Faculty of Mathematics, Physics and Informatics, Comenius University, Bratislava, Slovakia
\item \Idef{org40}Faculty of Nuclear Sciences and Physical Engineering, Czech Technical University in Prague, Prague, Czech Republic
\item \Idef{org41}Faculty of Science, P.J.~\v{S}af\'{a}rik University, Ko\v{s}ice, Slovakia
\item \Idef{org42}Faculty of Technology, Buskerud and Vestfold University College, Vestfold, Norway
\item \Idef{org43}Frankfurt Institute for Advanced Studies, Johann Wolfgang Goethe-Universit\"{a}t Frankfurt, Frankfurt, Germany
\item \Idef{org44}Gangneung-Wonju National University, Gangneung, South Korea
\item \Idef{org45}Gauhati University, Department of Physics, Guwahati, India
\item \Idef{org46}Helsinki Institute of Physics (HIP), Helsinki, Finland
\item \Idef{org47}Hiroshima University, Hiroshima, Japan
\item \Idef{org48}Indian Institute of Technology Bombay (IIT), Mumbai, India
\item \Idef{org49}Indian Institute of Technology Indore, Indore (IITI), India
\item \Idef{org50}Inha University, Incheon, South Korea
\item \Idef{org51}Institut de Physique Nucl\'eaire d'Orsay (IPNO), Universit\'e Paris-Sud, CNRS-IN2P3, Orsay, France
\item \Idef{org52}Institut f\"{u}r Informatik, Johann Wolfgang Goethe-Universit\"{a}t Frankfurt, Frankfurt, Germany
\item \Idef{org53}Institut f\"{u}r Kernphysik, Johann Wolfgang Goethe-Universit\"{a}t Frankfurt, Frankfurt, Germany
\item \Idef{org54}Institut f\"{u}r Kernphysik, Westf\"{a}lische Wilhelms-Universit\"{a}t M\"{u}nster, M\"{u}nster, Germany
\item \Idef{org55}Institut Pluridisciplinaire Hubert Curien (IPHC), Universit\'{e} de Strasbourg, CNRS-IN2P3, Strasbourg, France
\item \Idef{org56}Institute for Nuclear Research, Academy of Sciences, Moscow, Russia
\item \Idef{org57}Institute for Subatomic Physics of Utrecht University, Utrecht, Netherlands
\item \Idef{org58}Institute for Theoretical and Experimental Physics, Moscow, Russia
\item \Idef{org59}Institute of Experimental Physics, Slovak Academy of Sciences, Ko\v{s}ice, Slovakia
\item \Idef{org60}Institute of Physics, Academy of Sciences of the Czech Republic, Prague, Czech Republic
\item \Idef{org61}Institute of Physics, Bhubaneswar, India
\item \Idef{org62}Institute of Space Science (ISS), Bucharest, Romania
\item \Idef{org63}Instituto de Ciencias Nucleares, Universidad Nacional Aut\'{o}noma de M\'{e}xico, Mexico City, Mexico
\item \Idef{org64}Instituto de F\'{\i}sica, Universidad Nacional Aut\'{o}noma de M\'{e}xico, Mexico City, Mexico
\item \Idef{org65}iThemba LABS, National Research Foundation, Somerset West, South Africa
\item \Idef{org66}Joint Institute for Nuclear Research (JINR), Dubna, Russia
\item \Idef{org67}Konkuk University, Seoul, South Korea
\item \Idef{org68}Korea Institute of Science and Technology Information, Daejeon, South Korea
\item \Idef{org69}KTO Karatay University, Konya, Turkey
\item \Idef{org70}Laboratoire de Physique Corpusculaire (LPC), Clermont Universit\'{e}, Universit\'{e} Blaise Pascal, CNRS--IN2P3, Clermont-Ferrand, France
\item \Idef{org71}Laboratoire de Physique Subatomique et de Cosmologie, Universit\'{e} Grenoble-Alpes, CNRS-IN2P3, Grenoble, France
\item \Idef{org72}Laboratori Nazionali di Frascati, INFN, Frascati, Italy
\item \Idef{org73}Laboratori Nazionali di Legnaro, INFN, Legnaro, Italy
\item \Idef{org74}Lawrence Berkeley National Laboratory, Berkeley, California, United States
\item \Idef{org75}Lawrence Livermore National Laboratory, Livermore, California, United States
\item \Idef{org76}Moscow Engineering Physics Institute, Moscow, Russia
\item \Idef{org77}National Centre for Nuclear Studies, Warsaw, Poland
\item \Idef{org78}National Institute for Physics and Nuclear Engineering, Bucharest, Romania
\item \Idef{org79}National Institute of Science Education and Research, Bhubaneswar, India
\item \Idef{org80}Niels Bohr Institute, University of Copenhagen, Copenhagen, Denmark
\item \Idef{org81}Nikhef, Nationaal instituut voor subatomaire fysica, Amsterdam, Netherlands
\item \Idef{org82}Nuclear Physics Group, STFC Daresbury Laboratory, Daresbury, United Kingdom
\item \Idef{org83}Nuclear Physics Institute, Academy of Sciences of the Czech Republic, \v{R}e\v{z} u Prahy, Czech Republic
\item \Idef{org84}Oak Ridge National Laboratory, Oak Ridge, Tennessee, United States
\item \Idef{org85}Petersburg Nuclear Physics Institute, Gatchina, Russia
\item \Idef{org86}Physics Department, Creighton University, Omaha, Nebraska, United States
\item \Idef{org87}Physics Department, Panjab University, Chandigarh, India
\item \Idef{org88}Physics Department, University of Athens, Athens, Greece
\item \Idef{org89}Physics Department, University of Cape Town, Cape Town, South Africa
\item \Idef{org90}Physics Department, University of Jammu, Jammu, India
\item \Idef{org91}Physics Department, University of Rajasthan, Jaipur, India
\item \Idef{org92}Physik Department, Technische Universit\"{a}t M\"{u}nchen, Munich, Germany
\item \Idef{org93}Physikalisches Institut, Ruprecht-Karls-Universit\"{a}t Heidelberg, Heidelberg, Germany
\item \Idef{org94}Politecnico di Torino, Turin, Italy
\item \Idef{org95}Purdue University, West Lafayette, Indiana, United States
\item \Idef{org96}Pusan National University, Pusan, South Korea
\item \Idef{org97}Research Division and ExtreMe Matter Institute EMMI, GSI Helmholtzzentrum f\"ur Schwerionenforschung, Darmstadt, Germany
\item \Idef{org98}Rudjer Bo\v{s}kovi\'{c} Institute, Zagreb, Croatia
\item \Idef{org99}Russian Federal Nuclear Center (VNIIEF), Sarov, Russia
\item \Idef{org100}Russian Research Centre Kurchatov Institute, Moscow, Russia
\item \Idef{org101}Saha Institute of Nuclear Physics, Kolkata, India
\item \Idef{org102}School of Physics and Astronomy, University of Birmingham, Birmingham, United Kingdom
\item \Idef{org103}Secci\'{o}n F\'{\i}sica, Departamento de Ciencias, Pontificia Universidad Cat\'{o}lica del Per\'{u}, Lima, Peru
\item \Idef{org104}Sezione INFN, Bari, Italy
\item \Idef{org105}Sezione INFN, Bologna, Italy
\item \Idef{org106}Sezione INFN, Cagliari, Italy
\item \Idef{org107}Sezione INFN, Catania, Italy
\item \Idef{org108}Sezione INFN, Padova, Italy
\item \Idef{org109}Sezione INFN, Rome, Italy
\item \Idef{org110}Sezione INFN, Trieste, Italy
\item \Idef{org111}Sezione INFN, Turin, Italy
\item \Idef{org112}SSC IHEP of NRC Kurchatov institute, Protvino, Russia
\item \Idef{org113}SUBATECH, Ecole des Mines de Nantes, Universit\'{e} de Nantes, CNRS-IN2P3, Nantes, France
\item \Idef{org114}Suranaree University of Technology, Nakhon Ratchasima, Thailand
\item \Idef{org115}Technical University of Ko\v{s}ice, Ko\v{s}ice, Slovakia
\item \Idef{org116}Technical University of Split FESB, Split, Croatia
\item \Idef{org117}The Henryk Niewodniczanski Institute of Nuclear Physics, Polish Academy of Sciences, Cracow, Poland
\item \Idef{org118}The University of Texas at Austin, Physics Department, Austin, Texas, USA
\item \Idef{org119}Universidad Aut\'{o}noma de Sinaloa, Culiac\'{a}n, Mexico
\item \Idef{org120}Universidade de S\~{a}o Paulo (USP), S\~{a}o Paulo, Brazil
\item \Idef{org121}Universidade Estadual de Campinas (UNICAMP), Campinas, Brazil
\item \Idef{org122}University of Houston, Houston, Texas, United States
\item \Idef{org123}University of Jyv\"{a}skyl\"{a}, Jyv\"{a}skyl\"{a}, Finland
\item \Idef{org124}University of Liverpool, Liverpool, United Kingdom
\item \Idef{org125}University of Tennessee, Knoxville, Tennessee, United States
\item \Idef{org126}University of the Witwatersrand, Johannesburg, South Africa
\item \Idef{org127}University of Tokyo, Tokyo, Japan
\item \Idef{org128}University of Tsukuba, Tsukuba, Japan
\item \Idef{org129}University of Zagreb, Zagreb, Croatia
\item \Idef{org130}Universit\'{e} de Lyon, Universit\'{e} Lyon 1, CNRS/IN2P3, IPN-Lyon, Villeurbanne, France
\item \Idef{org131}V.~Fock Institute for Physics, St. Petersburg State University, St. Petersburg, Russia
\item \Idef{org132}Variable Energy Cyclotron Centre, Kolkata, India
\item \Idef{org133}Warsaw University of Technology, Warsaw, Poland
\item \Idef{org134}Wayne State University, Detroit, Michigan, United States
\item \Idef{org135}Wigner Research Centre for Physics, Hungarian Academy of Sciences, Budapest, Hungary
\item \Idef{org136}Yale University, New Haven, Connecticut, United States
\item \Idef{org137}Yonsei University, Seoul, South Korea
\item \Idef{org138}Zentrum f\"{u}r Technologietransfer und Telekommunikation (ZTT), Fachhochschule Worms, Worms, Germany
\end{Authlist}
\endgroup